\documentclass[aps,prb,english,amsmath,amssymb]{revtex4-1}
\usepackage[T1]{fontenc}
\usepackage[latin9]{inputenc}
\usepackage{graphicx}

\usepackage{babel}

\global\long\def\i{\imath}
\global\long\def\plll{\mathcal{P}_{\mathrm{LLL}}}
\global\long\def\plcs{\mathcal{P}_{\mathrm{LCS}}}
\global\long\def\elliptic#1#2#3{\vartheta_{#1}\!\left(\left.#2\vphantom{#3}\right|#3\right)}
\global\long\def\ket#1{\left|#1\right\rangle }
\global\long\def\bra#1{\left\langle #1\right|}
\global\long\def\braket#1#2{\left\langle #1\left|\vphantom{#1}#2\right.\right\rangle }
\global\long\def\ketbra#1#2{\left|#1\vphantom{#2}\right\rangle \left\langle \vphantom{#1}#2\right|}
\global\long\def\braOket#1#2#3{\left\langle #1\left|\vphantom{#1#3}#2\right|#3\right\rangle }
\global\long\def\ellipticgeneralized#1#2#3#4{\vartheta\left[\begin{array}{c}
      #1\\
      #2 
    \end{array}\right]\left(\left.#3\vphantom{#4}\right|#4\right)}

\begin{document}

\title{Coherent State Wave Functions on a Torus with a Constant Magnetic Field}

\author{M. Fremling}

\affiliation{Department of Physics,
Stockholm University \\
AlbaNova University Center\\
SE-106 91 Stockholm, Sweden}

\date{\today}

\begin{abstract}
We study two alternative definitions of localized states in the lowest Landau level (LLL) on a torus.
One definition is to construct localized states,
as projection of the coordinate delta function onto the LLL.
Another definition, proposed by Haldane,
is to consider the set of functions which have all their zeros at a single point.
Since a LLL wave function on a torus,
supporting $N_\phi$ magnetic flux quanta,
is uniquely defined by the position of its $N_\phi$ zeros,
this defines a set of functions that are expected to be localized 
around the point maximally far away form the zeros. 
These two families of localized states have many properties in common 
with the coherent states on the plane and on the sphere, 
{\em viz.} a resolution of unity and a self-reproducing kernel.
However, we show that  only the projected delta function is maximally localized.
Additionally, we show how to project onto the LLL,
functions that contain holomorphic derivatives and/or anti-holomorphic polynomials,
and apply our methods in the description of hierarchical quantum Hall liquids

\end{abstract}

\maketitle


\section{Introduction}

Fractional quantum Hall (QH) states are formed when a very clean 
two dimensional electron gas at low temperature 
is subjected to a strong magnetic field $B$\cite{Tsui_82}.
A defining feature of these liquid phases is a quantized value of the 
Hall conductance $\sigma_H=\nu e^2/\pi$, where $\nu$ is the filling fraction,
\emph{i.e.} the density of electrons in units of the density of a filled Landau level.
The topological nature of these states is also reflected in 
fractionally charged quasi particles\cite{Laughlin_83},
obeying fractional statistics\cite{Arovas_84},
protected edge modes, and a characteristic ground state degeneracy on higher genus manifolds\cite{Tau_86}.

At a semi-classical level, an electron in a strong magnetic field can be viewed
as a localized  distribution of charge with the size of the magnetic length 
$\ell_B=\sqrt{\hbar c/eB}$, and an orbital angular momentum $l$ 
associated with the cyclotron motion around the guiding center position $\mathbf{R}$.
For $\nu<1$ all the low energy states reside in the lowest Landau level (LLL),
where the maximally localized states have zero angular momentum with respect to the guiding center.
Technically, the wave functions of these localized electrons are coherent states with a Gaussian charge density profile.
It is rather natural to try to use these coherent states as a basis 
when trying to formulate a theory of the fractional QH effect; 
the cooperative ring exchange theory using coherent state functional integrals is an early example\cite{Kivelson_86}. 

With the success of the composite fermion wave functions describing the 
Jain series for $\nu=p/(2pm\pm1)$, and the idea of hierarchical states built 
from successive condensation of anyonic quasi particles, 
a significant effort has been put into finding ansatz wave functions directly in the position basis\cite{Haldane_83, Halperin_83, Jain_07_Book}.
The coherent states have rarely featured in such discussions,
although their usefulness for improving wave functions was stressed 
in an early paper by Girvin and Jach\cite{Girvin_84}. 

The construction of variational wave functions was put on a more solid 
theoretical ground by Moore and Read\cite{Moore_91} and Wen\cite{Wen_92} 
who proposed that representative QH wave functions for a large family 
of states could be constructed from conformal blocks in a conformal field theory (CFT).
This proposal was based on the deep connection between the Chern-Simons theories,
which give a low energy description of the QH liquids, and certain two-dimensional CFTs\cite{Witten_89}. 
The strength of this approach, compared with \emph{i.e.} the composite fermion description,
is that the topological properties are manifest, 
given certain well supported technical assumptions\cite{Bondersson_11}. 

In two recent papers\cite{Soursa_11a,Soursa_11b}
it is argued that the natural interpretation of the conformal block wave functions,
is as wave functions in a coherent state basis rather than in coordinate basis.
In the simplest cases the two interpretations are equivalent,
but the use of coherent states allows for a comprehensive understanding of the full QH hierarchy, 
including not only the prominently observed states in the Jain series,
but also more exotic ones corresponding to alternating condensations of quasi particles and quasi holes. 

Interesting topological properties are manifest in the QH wave functions on higher genus manifolds,
such as the torus, which has genus $g=1$.
Although several examples of QH wave functions on the torus are known,
including the Laughlin states and the non-abelian Moore-Read state,
there is no understanding of the hierarchy at the same level 
as on the plane\cite{Soursa_11a}, or on the sphere\cite{Kvorning_11}.
An important step towards this goal is to construct and analyze coherent states on the torus,
and this is the subject of this paper. 

In 1963 Glauber\cite{Glauber_63} pointed out that there is more than one way of defining a coherent state for the harmonic oscillator.
The most common definition is as an eigenstate of the annihilation operator $a$.
Equivalently, a coherent state can be defined as a state with minimal dispersion in phase space.
In the LLL, the analogy of the latter definition, is a state that minimizes the spatial dispersion
of the guiding center coordinates of the Landau orbit.
Alternatively, this can be viewed as the projection of a position eigenstate onto the LLL\cite{Malkin_70}. 

The structure of the paper is as follows:
On a torus with fixed boundary conditions,
the continuous group of magnetic translations that is present on the plane,
is broken to a group of finite translations.
This group defines a lattice, with $N_s^2$ number of points,
where $N_s$ is the number of states in the LLL.
In his original paper on QH wave functions on the torus,
Haldane proposed a set of coherent states that is naturally defined on 
this lattice\cite{Haldane_85}.
In Section \ref{sec:LCS} we study these states, 
which we will refer to as lattice coherent states (LCS). 
In Section \ref{sec:CCS} we construct a second set of coherent states 
in direct analogy to those discussed in the previous paragraph; 
\emph{i.e.} by projecting position eigenstates, 
which on the torus are periodic delta functions, onto the LLL.
Since these coherent states are labeled by the guiding center coordinate,
we will refer to them as continuous coherent states (CCS).
In Section \ref{sec: Localization} we  explore the localization properties of 
both the LCS and the CCS for tori described by an arbitrary modular parameter, $\tau$.

An important issue in formulating hierarchical QH wave functions on the torus,
is how to incorporate holomorphic derivatives, 
and this issue is discussed in Section \ref{sec:Derivatives-in-LLL}.
There we find that a holomorphic derivative projected on the LLL becomes a sum of translation operators.
We also discuss alternative ways of mapping higher Landau level functions onto the LLL.

Finally in Section \ref{sec:Laughlin4-1} we give an explicit example of how anti-holomorphic components can be treated,
by interpreting these functions in the basis of CCS.
We will in this Section find that the required ground state multiplicity is not manifest from the CFT construction,
but that it can be extracted by projection onto well defined momentum states.

For a review of coherent states see Ref. \onlinecite{Galetti_96} and 
\onlinecite{Zhang_90} and the comprehensive book by Perelomov\cite{Perelomov_86} for details.
Recently Kovalski \emph{et al.} has written several articles about different coherent 
states\cite{Kowalski_05,Kowalski_07,Kowalski_96} on the circle, sphere and plane.

In the following section we will begin by a short summary of some properties of charged particles in a magnetic field.


\section{Charged particles in a magnetic field on a torus}\label{sec:The-torus}

To fix the notation, we here summarize elementary facts about charged particles on a torus, 
and in a transverse magnetic field.
In particular, we give expressions for the LLL single-particle wave functions.

\subsection{Magnetic translation operators}

We use the Landau gauge $\mathbf{A}=By\hat{\mathbf{x}}$,
with the magnetic field  pointing in the $\hat{\mathbf{z}}$-direction,
$\mathbf{B}=B\hat{\mathbf{z}}$.
We use units where both the cyclotron frequency and the magnetic length are  set to unity, 
such that $\omega_c=eB/m=1$ and $\ell_B=\sqrt{\hbar/eB}=1$.
A general magnetic translation by a displacement $\mathbf L$,
can  be written in the Landau gauge as 
$t(\mathbf{L})=\exp[\mathbf{L}\cdot(\nabla+\imath y\hat{\mathbf{x}})-\imath\hat{\mathbf{z}}\cdot(\mathbf{L}\times\mathbf{r})]$.
We use the complex notation
$t(\alpha+\imath\beta)\equiv t(\alpha\hat{\mathbf{x}}+\beta\hat{\mathbf{y}})$,
and consider a torus spanned by the lattice vectors
$\mathbf{L}_1=L_x\hat{\mathbf{x}}$ and $\mathbf{L}_2=L_\Delta\hat{\mathbf{x}}+L_y\hat{\mathbf{y}}$.
Interpreting the torus as a parallelogram with the opposite sides identified,
$L_x$, $L_y$  are the width and height whereas $L_\Delta$ is the skewness.
For translations along the cycles of the torus we have 
\begin{eqnarray}
  t(\alpha)  &=&  e^{\alpha\partial_x} \nonumber \\
  t(\imath\beta) &=&  e^{\imath\beta x}e^{\beta\partial_y} \nonumber \\
  t(\alpha+\imath\beta)  &=&  e^{\imath\frac{1}{2}\alpha\beta} t(\imath\beta) t(\alpha).
\end{eqnarray}
Hence, magnetic translations around the cycles of the torus commute only if $L_xL_y=2\pi N_s$,
where $N_s$ is an integer.
The torus can thus be parametrized by two parameters: The complex modular parameter 
\begin{equation}
  \tau=\frac{1}{L_x}(L_\Delta+\imath L_y),
\end{equation}
and the number of magnetic flux quanta $N_s$. 
A general boundary condition on a wave function $\psi$ is given by 
\begin{eqnarray}
t(L_x)\psi(z) & = & e^{\imath\phi_1}\psi(z) \nonumber \\
t(\tau L_x)\psi(z) & = &e^{\imath\phi_2}\psi(z),\label{eq:bc}
\end{eqnarray}
where the phase angles $\phi_i$ have the physical interpretation of fluxes threading the two cycles of the torus.
We will use complex coordinates $z=x+\imath y$ although all wave functions will have a non-holomorphic Gaussian part.
The boundary conditions are not invariant under magnetic translations since
\begin{eqnarray*}
  t(L_x) t(\alpha+\imath\beta) \psi(z) & = & e^{\imath(\phi_1+\beta L_x)} t(\alpha+\imath\beta) \psi(z)\\
  & \vphantom{a}\\
  t(\tau L_x) t(\alpha+\imath\beta) \psi(z) & = & e^{\imath(\phi_2+\beta L_\Delta-\alpha L_y)} t(\alpha+\imath\beta) \psi(z),
\end{eqnarray*}
which follows from the magnetic translation operator algebra. 
Translating in one of the principal directions will change the boundary conditions along the conjugate principal axis.
From this we see that the boundary conditions are invariant under a subset of magnetic translations,
$\Gamma=\alpha + \imath\beta = \frac{2\pi n}{L_y}+\frac{2\pi m}{L_y}\tau=\frac{L_x}{N_s}n+\frac{L_x}{N_s}m\tau$
with integers $n$ and $m$. It is convenient to use the notation
\begin{eqnarray}
  x_n & = & \frac{L_x}{N_s}n=\frac{2\pi}{L_y}n\label{eq:x_n} \nonumber\\
  y_n & = & \frac{L_y}{N_s}n=\frac{2\pi}{L_x}n\label{eq:y_n} \ 
\end{eqnarray}
to parametrize the natural sub-lattice formed by these translations which preserve the boundary conditions.
Note that $x_ny_m=2\pi\frac{n\cdot m}{N_s}$, so that $e^{\imath x_ny_{N_s}}=1$.
We will call the displacement operators that move one $N_s$:th step in the principal direction 
\begin{eqnarray}
  t_1  &\equiv& t(x_1) \nonumber \\
  t_2 &\equiv& t(\tau x_1)
\end{eqnarray}
respectively. 
In the following we shall fix the boundary conditions to $\phi_1=\phi_2=0$,
but all results can trivially be extended to arbitrary $\phi_i$ using the magnetic translation operators. 


\subsection{Basis states }
The Hamiltonian in the Landau gauge, with units restored, is 
\begin{equation*}
  \hat H = \frac{1}{2m}p_y^2 + \frac{1}{2m}(p_x-\frac{eB}{c}y)^2.
\end{equation*}
A simple and instructive way to construct the torus wave functions
is to first consider a cylinder with the axis along the $y$-direction.
The simultaneous eigenfunctions of $t_1$ and $\hat H$, are easily obtained as
\begin{equation}
  \chi_{n,s}(z)  =  \frac{1}{\sqrt{L_x\sqrt{\pi}}}e^{-\imath y_sx} H_n(y-y_s) 
  e^{-\frac{1}{2}(y-y_s)^2}
  \label{eq:cylinder_functions}
\end{equation}
where $H_n$ is a Hermite polynomial, and $\hat{H}\chi_{n,s}=\hbar\omega_c(n+\frac{1}{2})\chi_{n,s}$.
The eigenvalue of $t_1$ is $t_1^l\chi_{n,s}=e^{-\imath y_lx_s}\chi_{n,s}$.
The states $\chi_{ns}$ can be obtained using ladder operators in
the form $a_s^{\dagger}=\frac{1}{\sqrt{2}}(y-y_s-\partial_y)$.
Using the form of \eqref{eq:cylinder_functions} one finds that $\partial_x\chi_{n,s}=-\imath y_s\chi_{n,s}.$
We can build ladder operators that are $s$-independent as 
$a^\dagger=\frac{1}{\sqrt{2}} (y-\imath 2 \partial_z)$
and $a=\frac{1}{\sqrt{2}}(y-\imath2\partial_{\bar{z}})$.
These operators are also independent of the choice of $\phi_1$, which $a_s^\dagger$ is not.
The operators $a^\dagger$ and $a$ are also the ladder operators on the torus.
Using $a$, $a^\dagger$ and $t_2$ we can, after fixing the fluxes $\phi_j$, generate the full set of torus basis wave functions.

We will however here make an explicit construction using $\chi_{n,s}$ and the required properties under $t_j$.
We want to build a linear combination $\eta_{n,s}=\sum_{m,r}a_{m,r}\chi_{m,r}$
with the properties:
$\hat H\eta_{n,s}=(n+\frac{1}{2})\eta_{n,s}$,
$t_1^l\eta_{n,s}=e^{-\imath y_lx_s}\eta_{n,s}$ and $t_2\eta_{n,s}=\eta_{n,s-1}$.
We must then choose $a_{m,r}=\delta_{m,n}\delta_{r,s}^{(N_s)}\exp(\imath\frac{1}{2}y_r\omega_r)$.
Here $\delta_{r,s}^{(N_s)}=\sum_t\delta_{r+N_st,s}$ is a periodic Kronecker delta,
and $\omega_r=\frac{L_\Delta}{N_s}r$ is the skewness of the $x_r\times\tau x_r$ lattice.
In the LLL, the basis functions can then be written as 
\begin{equation}
  \eta_s(z) = \frac{1}{\sqrt{L_x\sqrt{\pi}}}\sum_t
  e^{\imath\frac{1}{2}(y_s+tL_y)(\omega_s+tL_\Delta)}
  e^{-\imath(y_s+tL_y)x}
  e^{-\frac{1}{2}(y-y_s-tL_y)^2}
  \label{eq:Basis_Fourier}
\end{equation}
or in a more compact form as 
\begin{equation}
  \eta_s(z)=\frac{e^{-\frac{1}{2}(y-y_s)^2}}{\sqrt{L_x\sqrt{\pi}}}
  e^{-\imath y_sx}e^{\frac{1}{2} \imath\omega_s y_s}
  \elliptic 3{\frac{\pi N_s}{L_x}(z-\tau x_s)}{N_s\tau}.
  \label{eq:Basis_theta} 
\end{equation}
Here $\vartheta_3$ is the third quasi-periodic Jacobi theta function,
\begin{equation*}
  \elliptic 3z{\tau}=\sum_ke^{\imath\pi\tau k^2}e^{\imath2kz}.
  \end{equation*}
These functions are orthonormal on any translation of the fundamental domain, $L_x\times L_x\tau$.
In an orthogonal basis $\eta_s$ diagonal in $t_1$,
\emph{i.e.} $t_1\eta_s = e^{\imath x_s y_1} \eta_s$,
the operator $t_2$ will act as $t_2 \eta_s = \eta_{s+1}$ thus generating the full set of basis states. 
Since $t_1$ and $t_2$ do not commute we  may instead choose to diagonalize the $t_2$ operator and would get the eigenfunctions:
\begin{equation}
  \varphi_l(z)=\frac{e^{-\frac{1}{2}y^2}}{\sqrt{N_sL_x\sqrt{\pi}}}
  \elliptic 3{\frac{\pi}{L_x}(z+x_l)}{\frac{\tau}{N_s}}.
  \label{eq:var_psi_basis_final}
\end{equation}
The $t_2$ eigenfunction can also be obtained from $\eta_s$
by performing the modular transformation $\tau\rightarrow-\frac{1}{\tau}$
followed by letting $z\rightarrow\frac{|\tau|}{\tau}z$ and applying
the appropriate gauge transformation.


\section{Lattice coherent states}\label{sec:LCS}

All the states in the LLL are uniquely defined by the positions of the $N_s$ zeros in the wave function,
so we should be able to engineer a spatially localized state by choosing the position of these zeros appropriately.
Haldane and Rezayi \cite{Haldane_85} proposed a candidate for such a localized wave function,
obtained by putting all zeros in the same point.
Because the continuous translations of the plane have been broken down to a discrete set of translations,
generated on the torus by $t_1$ and $t_2$,
this point cannot be chosen arbitrarily.
Instead we will get a family of $N_s^2$ wave functions.
These  wave functions form an over-complete set, 
and we shall refer to them as lattice coherent states (LCS).
One usually thinks of coherent states as states that minimize the spatial dispersion.
As we will see in Section \ref{sec: Localization}, the LCS are, in this sense,
only coherent on a rectangular ($\Re(\tau)=0$) torus.
These states are interesting since they naturally incorporate the finite
translation structure of the translation operators that preserve the boundary conditions, $t_1$ and $t_2$.
To our knowledge this is the first time these states are carefully examined,
so we shall work out their properties, including in particular,
formulas for the resolution of unity,
and the reproducing kernel, in some detail. 

The most general wave function in the LLL on a torus is
\begin{equation}
  \psi(z)  =  \mathcal{N}e^{-\frac{y^2}{2}}e^{\imath kz}\prod_{j=1}^{N_s}
  \elliptic 1{\frac{\pi}{L_x}(z-\xi_j)}{\tau}
  \label{eq:Haldane_Rezaye}
\end{equation}
where $\xi_j$ is the position of the $j$:th zero.
Here $\vartheta_1$ is the first Jacobi theta function, 
\begin{equation*}
  \elliptic 1 z \tau  = 2\sum_{n=0}^{\infty}(-1)^{n}e^{\imath\pi\tau\left(n+\frac{1}{2}\right)^{2}}
  \sin\left(\left(2n+1\right)z\right).
  \end{equation*}
By demanding that $\psi(z)$ obeys periodic boundary conditions defined by 
\eqref{eq:bc}, we get relations on $k$ and 
$\bar{\xi}=\frac{1}{N_s}\sum_j\xi_j$.
These can be written as
\begin{eqnarray}
  e^{\imath kL_x}       & = & (-1)^{N_s} \nonumber\\
  e^{\imath \bar{\xi}L_y} & = & (-1)^{N_s}e^{-\imath kL_x\tau}.\label{eq:constraint}
\end{eqnarray}
Solving \eqref{eq:constraint} leads to the relations
\begin{eqnarray*}
  k             & = & \frac{L_y}{2}+y_n\\
  \bar{\xi}_x   & = & x_l-\frac{k}{L_y}(L_\Delta-L_x)\\
  \bar{\xi}_y   & = & -k.
\end{eqnarray*}
Here $n$ and $l$ are arbitrary integers.
We can thus write $\bar{\xi}$ as $\bar{\xi} = x_1[m-n\tau]-\frac{L_x}{2}[\tau-1]$, where $m=n+l$.
We define $z_j=\xi_j+\frac{1}{2}(1+\tau)L_x$; since we expect that the LCS,
where all the zeros $\xi_j$ are at the same point,
will have the maximum at the position diametrically opposed to $\bar{\xi}$.
These coordinates, $z_j$, will give possible positions of the maximum value.
Under this re-parametrization the general wave function can be written as 
\begin{equation}
  \psi(z) = \mathcal{N}e^{-\frac{y^2}{2}}e^{\imath(\frac{L_y}{2}+y_n)z}
  \prod_{j=1}^{N_s}\elliptic 1{\frac{\pi}{L_x}(z-z_j)-\frac{\pi}{2}(1-\tau)}{\tau}.
  \label{eq:haldane_new_coordinates}
\end{equation}
To proceed we will use the Jacobi theta function identities
\begin{eqnarray*}
  \elliptic 1 {z\pm\frac{\pi}{2}} \tau & = & \pm\elliptic 2 z \tau\\
  \elliptic 2 {z+\frac{\pi}{2}\tau} \tau & = & e^{-\imath\frac{\pi}{4}\tau}e^{-\imath z}\elliptic 3 z \tau
\end{eqnarray*}
to transform $\vartheta_1$ into $\vartheta_3$.
Here $\vartheta_2$ is the second Jacobi theta function
\begin{equation*}
  \elliptic 2 z \tau = 2\sum_{n=0}^{\infty}e^{\imath\pi\tau(n+\frac{1}{2})^{2}}\cos\left(\left(2n+1)\right)z\right).
\end{equation*}
We have up to a scale factor and a constant phase
\begin{equation}
  \psi(z) \propto e^{-\frac{y^2}{2}}e^{\imath y_nz}
  \prod_{j=1}^{N_s}\elliptic 3{\frac{\pi}{L_x}(z-z_j)}{\tau}.
  \label{eq:haldane_as_theta3}
\end{equation}
Defining the mean value of the zeros, $\bar{z}=\frac{1}{N_s}\sum_{j=1}^{N_s}z_j$ and assuming all $z_j=\bar{z}$ we get the requirement $\bar{z}=x_m-x_n\tau$,
where again  $n$ and $m$ are integers.
Changing $n\rightarrow-n$ we get the full wave function,
with some $n$ and $m$ specific normalization $\mathcal{N}_{nm}$, to be
\begin{equation}
  \psi_{nm}(z) = \mathcal{N}_{nm}e^{-\frac{y^2}{2}}e^{-\imath y_nz}
  \elliptic 3{\frac{\pi}{L_x}(z-\bar{z}_{nm})}{\tau}^{N_s}
  \label{eq:LCS_theta_product}
\end{equation}
where $\bar z_{nm}=x_m+x_n\tau$.
For numerical evaluation, equation \eqref{eq:LCS_theta_product} is a useful expression.
However, for analytic manipulations this is not the best way of writing $\psi_{nm}(z)$,
and it also leaves unanswered the question of how to calculate the normalization $\mathcal{N}_{nm}$.
We solve for the relative normalization by using the magnetic translation operators to transform $\psi_{00}$ into $\psi_{nm}$.
We see by inspection that $|\mathcal{N}_{nm}|=\mathcal{N}e^{-\frac{y_n^2}{2}}$,
where $\mathcal{N}\equiv\mathcal{N}_{00}$.


\subsection{Fourier expanding $\psi_{nm}$ }

To bring $\psi_{nm}$ to a form that facilitates further manipulations we expand $\vartheta_3^{N_s}$ in Fourier modes as
\begin{eqnarray*}
  \elliptic 3{\frac{\pi}{L_x}z}\tau^{N_s} 
  & = & \sum_{\{ k_j\} =-\infty}^{\infty}
  e^{\imath\pi\tau\sum_{j=1}^{N_s}k_j^2}e^{2\imath\frac{\pi}{L_x}z\sum_{j=1}^{N_s}k_j}\\
  & = & \sum_{K=-\infty}^{\infty} e^{\frac{\imath}{2}\tau x_Ky_K}e^{\imath y_Kz}Z_K.
\end{eqnarray*}
In the second line we have made the substitution $k_j=\frac{K}{N_s}+\tilde{k}_j$, where $K=\sum_{j=1}^{N_s}k_j$, 
such that $\tilde{k}_j$ is the deviation from the mean value of $k_j$.
 We hide the nontrivial sum over $\tilde{k}$ in the value $Z_K$, 
\begin{equation}
  Z_K=\sum_{\stackrel{\{ k_j\} =-\infty}{\sum_{j=1}^{N_s}k_j=K}}^{\infty}e^{\imath\pi\tau\sum_{j=1}^{N_s}\tilde{k}_j^2}.
  \label{eq:Z_K}
\end{equation}
The constant $Z_K$ can, together with the factor $e^{-\imath\pi\tau\frac{K^2}{N_s}}$,
be interpreted as the partition function of $N_s$ particles
on a circle with the total angular momentum $K$.
The expression for $\psi_{00}$ is thus 
\begin{equation*}
  \psi_{00}(z) = \mathcal{N}\sum_{K=-\infty}^\infty Z_Ke^{-\frac{1}{2}(y+y_K)^2}e^{\imath y_K(x+\frac{\omega_K}{2})}
\end{equation*}
which for general $\psi_{nm}$ becomes 
\begin{equation}
  \psi_{nm}(z)=\mathcal{N}\sum_{K=-\infty}^\infty Z_{K+n}e^{-\frac{1}{2}(y+y_K)^2}e^{\imath y_K(x-x_m)}e^{\imath\frac{1}{2}y_K\omega_K}.
  \label{eq:LCS_Fourier}
\end{equation}
We note that for equations \eqref{eq:LCS_Fourier}  and \eqref{eq:LCS_theta_product} to match,
we must have $\mathcal N_{nm}=e^{\imath y_n (x_m + \frac{1}{2} x_n \tau)}\mathcal N$.
The price for this simple formula is that we now have a set of complicated constants $Z_K$.
Fortunately for our purposes it is sufficient to know the periodicity property $Z_{K+N_s}=Z_K$.
It will turn out that the details about $Z_K$ can be hidden in the normalization.
$Z_K$ is periodic because it only contains deviations from the sum mean value $\frac{K}{N_s}$.


\subsection{Overlap and normalization}

The overlap between the two states $\psi_{nm}$ and $\psi_{n^{\prime}m^{\prime}}$
can be calculated by noting that the $x$-integration gives Kronecker deltas,
that allow us to combine the  $y$-integral from an incomplete to a complete Gaussian integral.
We decompose the sum over $K$ as $K=l+tN_s$, such that $\sum_{K=-\infty}^\infty = \sum_{l=1}^{N_s} \sum_{t=-\infty}^\infty $.
The overlap on any fundamental region is thus 
\begin{equation*}
  \braket{\psi_{n^\prime m^\prime}}{\psi_{nm}}
  = L_x\mathcal{N}^2 \sum_{l=1}^{N_s} Z_{l+n}Z_{l+n^\prime}^{\star}
  e^{\imath y_l(x_{m^\prime}-x_m)} \sum_{t=-\infty}^\infty
  \int_0^{L_y}dy\,e^{-(y+y_l+tL_y)^2}.
\end{equation*}
After the Gaussian integral is performed the overlap reduces to
\begin{equation}
  \braket{\psi_{n^\prime m^\prime }}{\psi_{nm}}
  =\sqrt\pi L_x\mathcal{N}^2
  \sum_{l=1}^{N_s}Z_{l+n}Z_{l+n^\prime}^{\star} e^{\imath y_l(x_{m^\prime}-x_m)}.
  \label{eq:overlap_nm_nm_prime}
\end{equation}
Choosing $m^{\prime}=m$ and $n^{\prime}=n$ we get 
\begin{equation}
  \sum_{l=1}^{N_s}|Z_l|^2=\frac{\mathcal{N}^{-2}}{L_x\sqrt{\pi}},
  \label{eq:normalization}
\end{equation}
which defines the normalization constant in equation \eqref{eq:LCS_Fourier}.


\subsection{Resolution of unity\label{sub:Resolution-of-unity}}
Just as the usual coherent states on the plane, 
the states  $\psi_{nm}$  allow for a simple resolution of unity in the LLL given as
\begin{equation} 
  \plll=\frac{1}{N_s}\sum_{m,n=1}^{N_s}\ketbra{\psi_{nm}}{\psi_{nm}}.
  \label{eq:LCS_r-o-u}
\end{equation} 
We now prove this relation, which will be useful in applications of the coherent states.
First notice that from \eqref{eq:LCS_Fourier} we have 
\begin{eqnarray}
  t_1^{m^{\prime}}t_2^{n^{\prime}}\ket{\psi_{nm}} & = & e^{\imath y_{n^\prime}x_m}\ket{\psi_{n-n^\prime,m-m^\prime}}\label{eq:t_1t_2_nm_v1}\\
  \bra{\psi_{nm}}t_1^{m^{\prime}}t_2^{n^{\prime}} & = & e^{\imath y_{n^\prime}(x_m+x_{m^\prime})}\bra{\psi_{n+n^\prime,m+m^\prime}},\label{eq:t_1t_2_nm_v2}
\end{eqnarray}
and from this, it follows that
\begin{equation*}
\left[t_2^{m^\prime}t_1^{n^\prime}\,,\,\sum_{m,n}\ketbra{\psi_{nm}}{\psi_{nm}}\right]=0,
\end{equation*}
by changing the summation index.
Since $\sum_{m,n}\ketbra{\psi_{nm}}{\psi_{nm}}$ commutes with any translation it must be proportional to the identity,
\emph{i.e.} $\sum_{m,n}\ketbra{\psi_{nm}}{\psi_{nm}}=c\,\plll$ where $c$ is a constant.
To determine this constant, it is sufficient to calculate a single matrix element.
To do this we first express $\plll$ in the complete basis $\ket{\eta_s}$ given by the wave functions \eqref{eq:Basis_theta},
such that
\begin{equation*}
  \braOket {z=0}{\plll}{z=0} =\sum_s|\braket {z=0}{\eta_s}|^2= 
  \frac{1}{L_x\sqrt{\pi}}\sum_{r,t}^{\infty}e^{-\i\omega_rL_yt+\frac{1}{2}\imath L_yL_\Delta^2}
  e^{-\frac{1}{2}(y_r-tL_y)^2-\frac{1}{2}y_r^2}.
\end{equation*}
Calculating the same matrix element using the lattice coherent states, we obtain
\begin{equation*}
  \sum_{mn}|\braket {z=0}{\psi_{nm}}|^2 = N_s\cdot\braOket {z=0}\plll {z=0}.
\end{equation*}
This implies that $c=N_s$ and thus proves \eqref{eq:LCS_r-o-u}.


\subsection{Self-reproducing kernel}\label{sec:Reproducing-kernel}

Since the, un-normalized, coherent states $\varphi_w(z)$ on the plane
can be obtained by projecting a position eigenstate onto the LLL,
they are closely related to the holomorphic delta function.
The state $\varphi_w$ is in fact a self-reproducing kernel on the space of LLL wave functions,
\emph{i.e.} $\int d^2 w\, \varphi_w (z) \psi (w) = \psi(z)$.
If we instead convolute the kernel  $\varphi_w(z)$ with a wave function that
has components in higher Landau levels, these will simply be projected out. 
It is not obvious that there is a similar kernel for the LCS $\psi_{nm}$,
but we can in fact prove that 
\begin{equation}
  \psi(z)=S^{-1}\sum_{m,n=1}^{N_s}e^{\imath\frac{1}{2}y_n\omega_n}\psi_{nm}(z)\psi(z_{nm}),
  \label{eq:LCS_self_rep_kernel}
\end{equation}
where $\psi(z)$ is again an arbitrary wave function in the LLL and $z_{nm}=x_m+x_n\tau$.
If we substitute $\psi(z)$ with an arbitrary wave function $\phi(z,z^\star)$ then
\eqref{eq:LCS_self_rep_kernel}  defines a map, $\plcs$,
from the space of all wave functions to the LLL wave functions.
Although this map is {\em not} a projection on the LLL,
it might still be useful in constructing quantum Hall wave functions,
and we shall comment on this in Section \ref{sec:Derivatives-in-LLL}.

We now prove \eqref{eq:LCS_self_rep_kernel}  by first looking at the specific case where $\psi(z)$ is a basis wave function.
With a  bit of algebra we can establish 
\begin{equation*}
  \sum_{m,n=1}^{N_s}e^{\imath\frac{1}{2}y_n\omega_n}\psi_{nm}(z_{lp})\eta_s(z_{nm}) 
  = \eta_s(z_{lp})N_s\psi_{00}(0),
\end{equation*}
where $\eta_s$ is any basis wave function defined in \eqref{eq:Basis_Fourier}.
From this follows
\begin{equation}
  \psi(z_{lk})=S^{-1}\sum_{m,n=1}^{N_s}e^{\imath\frac{1}{2}y_n\omega_n}
  \psi_{nm}(z_{lk})\psi(z_{nm}),
  \label{eq:psi_s_rep_kernel}
\end{equation}
which holds for all $l,k\in\mathbb{Z}$ where $\psi$ is any state in the LLL,
and $S=N_s\psi_{00}(0)$ is a normalization factor.
That \eqref{eq:psi_s_rep_kernel} holds for $N_s\times N_s$ points is more than enough 
to ensure that \eqref{eq:LCS_self_rep_kernel} holds for arbitrary $z$.
To prove this, consider the difference between the left hand side and right
hand side of \eqref{eq:psi_s_rep_kernel}, for arbitrary $z$.
This difference, $\xi(z)$, must have zeros at all $z=z_{nm}$.
But $\xi(z)$ can fulfill the boundary conditions,
if and only if it has the form of \eqref{eq:Haldane_Rezaye},
which only has $N_s$ zeros.
Therefore $\xi(z)=0$ identically, which concludes the proof.

This LCS-map should be used with care,
because it is \emph{not} a projection onto the LLL.
Using $\plcs$ on non-LLL function will in general give nonzero results.
This is seen by considering $\delta(z-z^\prime)$ which has components in all Landau levels.
It is obvious that $\plcs\delta(z-z^{\prime})$ will give zero even though we know that $\delta(z-z^{\prime})$ has components in the LLL.
Thus the effect of $\plcs$ is that the contributions from non-LLL states precisely cancels the LLL part,
except at $z^{\prime}=z_{nm}$ for which the contribution is divergent.

It is instructive to consider $\plcs$ for the special case of $N_s=1$.
In this case there is only  one state and $\plcs\phi(z)=\frac{\psi_{00}(z)}{\psi_{00}(0)}\phi(0)$.
If $\phi$ is a basis state $\eta_{n,0}(z)$ we can, via a simple parity argument,
show that $\eta_{2n+1,0}(0)=0$ and $\eta_{2n,0}(0)\neq 0$ in general.
For $N_s\neq1$ a similar analysis is more difficult, 
since $\eta_{n,s}$ should be evaluated analytically in $N_s\times N_s$ points together with $\psi_{nm}(z_{l,p})$.
Numerical studies suggest that $\plcs\eta_{2n+1,s}=0$ and $\plcs\eta_{2n,s}\neq 0$ does hold for general $N_s$.
This would mean that if $\phi$ is restricted to the two lowest Landau levels, then $\plcs\phi=\plll\phi$.


\section{Continuous coherent states}\label{sec:CCS}

In the previous section we introduced the LCS wave functions as candidates for coherent states.
From the point of view of describing localized states,
the LCS have the drawback of only defining states that are localized around the finite set 
of points spanned by the lattice vectors $x_1$ and $\tau x_1$.
It is complicated to construct a localized particle around some other point with the LCS.
This brings us to the notion of continuous coherent states (CCS) 
that are obtained as the projection of a position eigenstate on the lowest Landau level.
We thus define 
\begin{equation}
  \varphi_{w}(z) = \plll \delta(z-w).
\end{equation}
Here $w=x^\prime+\imath y^\prime$.
As a projector we can either use some basis states 
$\plll=\sum_s\ketbra{\eta_s}{\eta_s}$ or we can use 
\eqref{eq:LCS_r-o-u} from Section \ref{sec:LCS}.
The overlap between two coherent states $\varphi_w$ and $\varphi_u$ is readily obtained as
\begin{equation}
  \braket u w =\sum_s\eta_s(u)\eta^\star_s(w)=\varphi_w(u).
  \label{eq:CCS_normalization}
\end{equation}
From the definition of $\varphi_w(z)$ also follows a resolution of unity 
\begin{equation*}
  \psi(z)=\int d^2w\varphi_w(z)\psi(w)
\end{equation*}
for states in LLL and zero otherwise.
For analytical manipulations we can either work with \eqref{eq:LCS_r-o-u}, and construct the LLL CCS as 
\begin{equation*}
  \varphi_w(z)=\frac{1}{N_s}\sum_{mn}\psi_{nm}(z)\psi_{nm}^{\star}(w),
\end{equation*}
or work with $\eta_s(z)$ and obtain $\varphi_w(z)$ using
 $\varphi_w(z)=\sum_{s=1}^{N_s}\eta_{s}^{\star}(w)\eta_{s}(z)$.
We can note that when we work with arbitrary boundary conditions, given by \eqref{eq:bc},
we can use the transform 
\begin{equation*}
  \varphi_w(z)\rightarrow t^{(z)}(\gamma)\cdot\varphi_w(z)\cdot t^{(w)}(-\gamma)
\end{equation*}
where $\gamma=\frac{x_1}{2\pi}(\phi_2+\phi_1\tau)$ and $t^{(z)}$ acts on the $z$ coordinate.
After some algebra, where we use the sums over $m$ and $n$ to cancel the $Z_K$ factors against the normalization, we get
\begin{equation}
  \varphi_w(z)  =  \frac{1}{L_x\sqrt{\pi}}\sum_{K,t}
  e^{-\frac{1}{2}(y+y_K)^2}
  e^{-\frac{1}{2}(y^{\prime}+y_K+L_yt)^2}
  e^{-\imath y_K(x^{\prime}-x)}
  e^{-\imath(\omega_K+x^{\prime})L_yt}
  e^{-\imath\frac{1}{2}L_yL_\Delta t^2}.
  \label{eq:CCS-expanded}
\end{equation}
With the aim of getting an expression in terms of $\vartheta$-functions,
we rewrite $\varphi_w$ as, 
\begin{eqnarray}
  \varphi_w(z) & = & P(y,y^{\prime})\times Y(z,w^\star) ,  \label{eq:CCS_prodct_P_Y}
\end{eqnarray}
which factorizes into a Gaussian part 
\begin{equation*}
  P(y,y^{\prime})=\frac{e^{-\frac{1}{2}(y^2+y^{\prime2})}}{L_x\sqrt{\pi}}
\end{equation*}
and a holomorphic part 
\begin{equation}
  Y(z, w^\star )  =  \sum_te^{-\imath w^\star tL_y}e^{-\imath\pi N_s\bar{\tau}t^2}E_t,
  \label{eq:Y(z,z)}
\end{equation}
where $E_t$ can be simplified as 
\begin{equation}
  E_t  =  \elliptic 3{\frac{\pi}{L_x}( w^\star -z)+\pi t\bar{\tau}}{\frac{2}{N_s}\imath\Im(\tau)}.
\label{eq:E_t(z,z)}
\end{equation}
In order to have an expression for $Y(z,w^\star)$ that is compact,
we would like to find a transformation that removes the term $\pi t\bar{\tau}$ from the $\vartheta_3$-function.
However since $\vartheta_3$ has only $\Im(\tau)$ as second argument this is not possible.
In the rectangular case where $\Re(\tau)=0$ we can proceed. 


\subsection{Purely imaginary $\tau$}

\begin{figure}
  \begin{centering}
    \includegraphics[width=0.3\columnwidth]{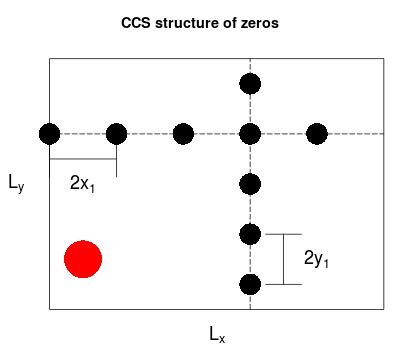}
    \includegraphics[width=0.3\columnwidth]{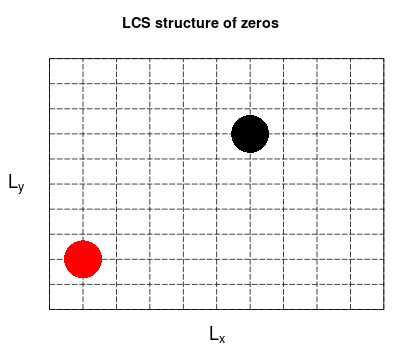}
    \par\end{centering}
    \caption{
      Structure of zeros of continuous coherent states $\varphi_w$ (left) and lattice coherent states $\psi_{nm}$ (right).
      The red circles represent the maximum density,
      whereas the black circles represent zeros.
      On the rectangular ($\Re(\tau)=0$) torus for $N_s$ even, the zeros of the CCS form a ``cross'' 
      centered over $z=w+\frac{L_x}{2}(1+\tau)$.
      For general $\tau$ and/or odd $N_s$ the structure of the zeros is not as easy to describe.
      \label{fig:zeros_CCS_LCS} }
\end{figure}

In the following we will assume that $\tau$ is purely imaginary \emph{i.e.} $\tau=\imath\frac{L_y}{L_x}$.
Even so, we will obtain different functional forms depending on whether $N_s$ is even or odd.
This is related to the structure of the zeros.
For an even $N_s$ the zeros will be divided into two groups.
Each group of zeros will, for any $w$, be regularly spaced on some line parallel to one of the fundamental axes.
For an odd $N_s$ the zeros cannot be divided into two grops, and the distributions of the zeros is more intricate.
For even $N_s$ we will have \eqref{eq:CCS_imag_tau_even_Ns} and for odd $N_s$ we will have \eqref{eq:CCS_imag_tau_odd_Ns}.
It will be convenient to define:
\begin{eqnarray}
  T^- & = & \frac{\pi}{L_x}(z-w^\star)\nonumber\\
  T^+ & = & \frac{\pi}{L_x}(z+w^\star)\label{eq:T^+}.
\end{eqnarray}
In the case of an even $N_s$ we can use the $\vartheta$-function relation
\begin{equation*}
  \elliptic 3 {z+k\tau\pi} \tau = e^{-\imath\pi\tau k^2}e^{-2\imath kz}\elliptic 3 z \tau
\end{equation*}
which reduces \eqref{eq:Y(z,z)} to 
\begin{equation*}
Y(z, w^\star )  =  \elliptic 3{T^-}{\frac{2\tau}{N_s}}
\elliptic 3{\frac{N_s}{2}T^+}{\frac{N_s\tau}{2}}.
\end{equation*}
The full form of the continuous coherent state for even $N_s$ is thus
\begin{equation}
  \varphi_w(z) = \frac{e^{-\frac{1}{2}(y^2+y^{\prime2})}}{L_x\sqrt{\pi}}
  \elliptic 3{T^-}{\frac{2\tau}{N_s}}
  \elliptic 3{\frac{N_s}{2}T^+}{\frac{N_s\tau}{2}}.
  \label{eq:CCS_imag_tau_even_Ns}
\end{equation}
We can see that the zeros of this coherent state lie on two lines intersecting at 
$z=w+\frac{1}{2}(1+\tau)L_x$ as drawn in the left panel of Figure \ref{fig:zeros_CCS_LCS}.
This structure is entirely different from the LCS,
where all the zeros are at the same point,
as indicated in the right panel of Figure \ref{fig:zeros_CCS_LCS}.

For odd $N_s$, the expansion of $E_t$ in \eqref{eq:E_t(z,z)} depends on whether $t$ is even or odd.
If $t$ is even we basically have the same result as in the case where $N_s$ is even 
since we know that $\frac{tN_s}{2}$ is an integer.
For the odd $t$-values we will need to use some more $\vartheta$-function relations 
\begin{eqnarray*}
  \elliptic 3 {z\pm\frac{1}{2}\pi\tau} \tau & = & q^{-\frac{1}{4}}e^{\mp\imath z}\elliptic 2 z \tau
  \label{eq:theta_3_half_tau_period}\\
  \elliptic 2 {z+n\pi\tau} \tau & = & q^{-n^2}e^{-2\imath nz}\elliptic 2 z \tau
  \label{eq:theta_2_tau_period}
\end{eqnarray*}
such that $\vartheta_3$ is transformed into $\vartheta_2$.
The final result for $\varphi_w(z)$ in the case where $N_s$ is odd is
\begin{equation}
  \varphi_w(z)  = \frac{e^{-\frac{1}{2}(y^2+y^{\prime2})}}{L_x\sqrt{\pi}}\sum_{j=2,3}
  \elliptic j{T^-}{\frac{2\tau}{N_s}}
  \elliptic j{T^+N_s}{2N_s\tau}.
  \label{eq:CCS_imag_tau_odd_Ns}
\end{equation}

\subsection{Generic $\tau$ results}

In the generic case we see that we cannot simplify $\varphi_w$ to a finite sum of $\vartheta$-functions.
We get for even $N_s$, after shifting away the $\pi t\Im(\bar{\tau})$ part, 
\begin{equation}
  \varphi_w(r) = \frac{e^{-\frac{1}{2}y^2-\frac{1}{2}y^{\prime2}}}{L_x\sqrt{\pi}}
  \sum_t e^{-\imath tN_sT^+} e^{-\frac{\pi N_s}{2}\Im(\tau)t^2}
  \elliptic 3{T^--\pi t\Re(\tau)}{\frac{2}{N_s}\imath\Im(\tau)}.
  \label{eq:CCS_generic_Ns_even}
\end{equation}
We see that if it was not for the term $\pi t\Re(\tau)$, 
in the $\vartheta$-function,
the projection would be given by \eqref{eq:CCS_imag_tau_even_Ns} albeit with $\tau\rightarrow\imath\Im(\tau)$.
For odd $N_s$ we get a similar result resembling \eqref{eq:CCS_imag_tau_odd_Ns}\footnote{
  Knowing the result in the square case, one could have guessed a generalization,
  $\varphi_w^{(\mathrm{guess})}(z)$,
  which is the solution for $\Re(\tau)=0$ but with $\tau$ allowed to be complex.
  That this form is incorrect, one can see from the modular behavior.
  For $\tau\rightarrow\tau+n$, where $n$ is not a multiple of $\frac{N_s}{2}$,
  stray zeros around $z=w$ can destroy the coherence of the state $\varphi_w^{(\mathrm{guess})}(z)$.
}.


\section{Localization behavior of CCS and LCS}\label{sec: Localization}
\begin{figure}
  \begin{tabular}{cccc}
    \includegraphics[width=0.2\columnwidth]{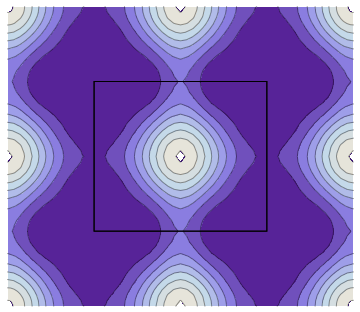} &
    \includegraphics[width=0.2\columnwidth]{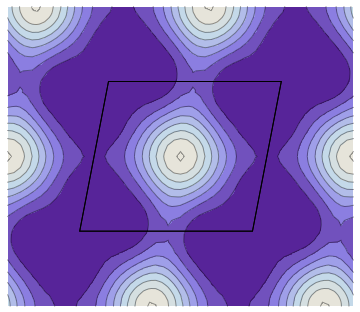} &
    \includegraphics[width=0.2\columnwidth]{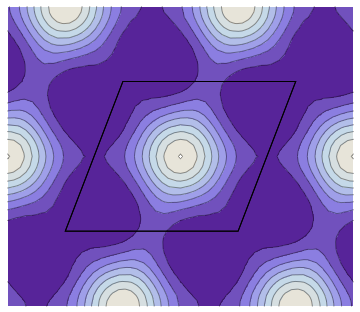} &
    \includegraphics[width=0.2\columnwidth]{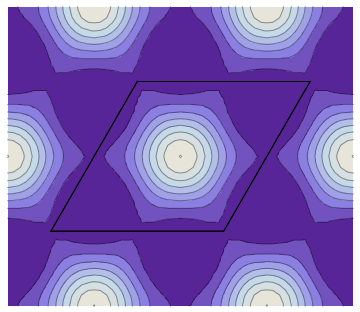}
    \tabularnewline
    $a$ & $b$ & $c$ & $d$
    \tabularnewline
    \includegraphics[width=0.2\columnwidth]{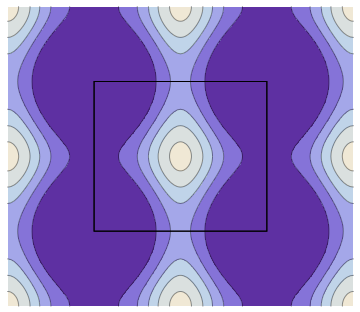} &
    \includegraphics[width=0.2\columnwidth]{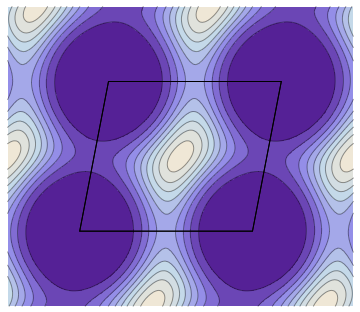} &
    \includegraphics[width=0.2\columnwidth]{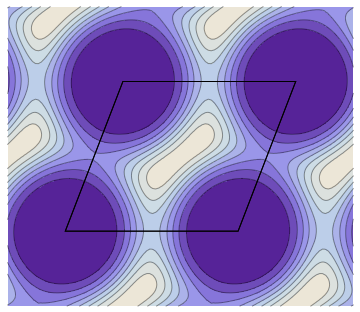} &
    \includegraphics[width=0.2\columnwidth]{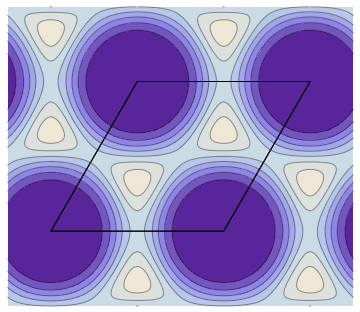}
  \end{tabular}
  \caption{
    The spatial profile of CCS (upper) and LCS (lower) at $N_s=4$.
    No axes are drawn as the black square shows the fundamental domain, $\Omega$.
    The columns correspond to $\tau=\sqrt{\frac{3}{4}}\imath+\tau_x$ where 
    a) $\tau_x=0$,
    b) $\tau_x=\frac 1 6$,
    c) $\tau_x=\frac 1 3$ and
    d) $\tau_x=\frac 1 2$.
    The values of $\tau_x$ correspond to a rectangular,
    two general and one triangular lattice.
    Lighter colours correspond to larger values of $|\psi|^2$.
    The CCS nicely reshapes itself whereas the LCS becomes very distorted
    and develops a double maxima.
    The double maxima should be stable even for large values of $N_s$ because of symmetry.
    \label{fig:spatial_profile}}
\end{figure}

In this section we will analyze the localization properties of the LCS and the CCS wave functions 
by calculating $\sigma_x\sigma_y$ for different $\tau$ and $N_s$,
where $\sigma_x^2= \langle x^2\rangle -\langle x\rangle^2$.
We will for simplicity consider projections of $w=0$ unless stated otherwise.
We will calculate the spatial dispersion using 
$\langle A(x,y)\rangle = \int dx\,dy\, A(x,y)\,|\varphi_w(z)|^2$,
where the integral runs over a fundamental domain $\Omega$, chosen so
that $\langle x\rangle$ and $\langle y\rangle$ are at the center of $\Omega$.
For highly skew tori the LCS might not be localized around only a single point.
For instance, we will see that if we set $\tau=\imath\frac {\sqrt 3} 2  +\frac 1 2$,
which corresponds to a triangular lattice,
then $\psi_{nm}(z)$ develops two maxima.
These will by symmetry be located at the center of each triangle and form a honeycomb lattice,
as can be seen in Figure \ref{fig:spatial_profile}.
This feature should be stable even for $N_s\rightarrow\infty$,
meaning that the LCS can never describe a localized state if $\tau$ is tilted enough.

Figure \ref{fig:spatial_profile} shows how the spatial profile changes as $\tau$ is tuned away from $\Re(\tau)=0$.
The CCS retain their circular shapes even at small $N_s$.
For the same change in $\tau$, the LCS become very distorted\footnote{
  There is a subtle point that should be made here to avoid confusion.
  For odd $N_s$, if one lets $\tau\rightarrow\tau+1$,
  it looks like the CCS get distorted.
  This effect is because as we tilt the lattice from $\tau$ to $\tau+1$ 
  the $w$ coordinate of the CCS will move from $w=x_n+\tau x_m$ to 
  $w=x_{n+\frac{1}{2}}+\tau x_{m+\frac{1}{2}}$ causing the distortion.}.
The LCS distortion is a combination of a geometric effect, and that the LCS develop multiple maxima.
The geometric effect is because $\Omega_\tau$ has its center at the boundary of $\Omega_{\tau+1}$, if they share one corner.
Because of this effect we need to be careful with the measure $\sigma_x\sigma_y$.
It is not guaranteed that the maximum of $|\psi|^2$ is located at $z=\langle x\rangle +\imath\langle y\rangle$.

\begin{figure}
  \begin{tabular}{cc}
    \includegraphics[width=0.3\columnwidth]{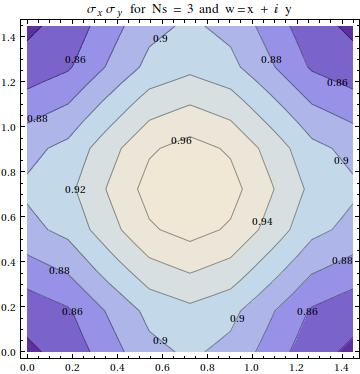} &
    \includegraphics[width=0.4\columnwidth]{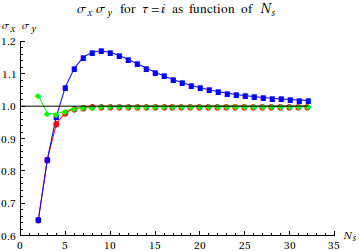} 
    \tabularnewline
    $a$ & $b$
  \end{tabular}
  \caption{
    The spatial dispersion for $\tau=\imath$ measured as $\sigma_x\sigma_y$.
    Left:
    $\sigma_x\sigma_y$ for $\varphi_w$ with variation of $w$.
    The smallest dispersion is at $w=0$ whereas the largest is at $w=\frac 1 2(1+\tau)$.
    Right:
    $\sigma_x\sigma_y$ for a CCS, (red) on $w=0$ and (green) on $w=\frac 1 2(1+\tau)$, 
    and LCS (blue) for some values of $N_s$.
    The CCS in general have smaller dispersion than the LCS 
    but the dispersion of CCS depends on $w$.
    For $N_s=1,2,3$ the value of $\sigma_x\sigma_y$ coincides for LCS and CCS because 
    the wave functions are the same these cases.
    \label{fig:dxdx_of_Ns}}
\end{figure}

Figure \ref{fig:dxdx_of_Ns} shows how the dispersion depends on the number of fluxes, $N_s$.
The dispersion is significantly smaller for small values of $N_s<10$,
because at these sizes the torus is so small that the CCS have the same width as the torus.
At $N_s>10$ the CCS have saturated at the dispersion value expected on the plane,
whereas it takes the LCS to $N_s>40$ to reach the same values.
We can get a feeling for why the CCS are more localized than the LCS by noticing
the structure of zeros for the CCS and LCS.
It is likely that as the LCS have all their zeros diametrically opposed to the maximum value 
they do not localize the wave function as efficient as the CCS do 
where the zeros are spread along the border of the torus.

\begin{figure}
  \begin{tabular}{ccc}
    \includegraphics[width=0.2\columnwidth]{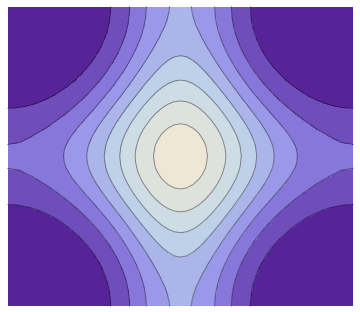} &
    \includegraphics[width=0.2\columnwidth]{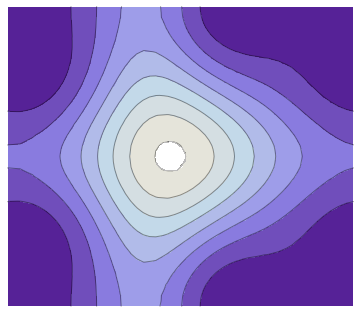} &
    \includegraphics[width=0.2\columnwidth]{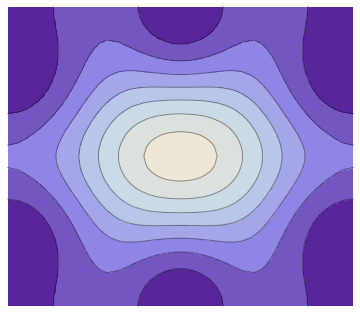} 
    \tabularnewline
    $a$ & $b$ & $c$
    \tabularnewline
    \includegraphics[width=0.2\columnwidth]{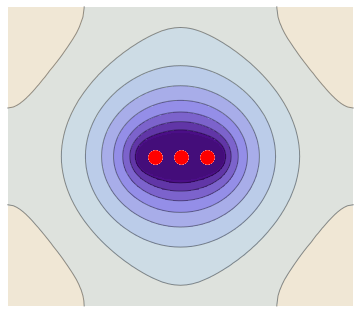} &
    \includegraphics[width=0.2\columnwidth]{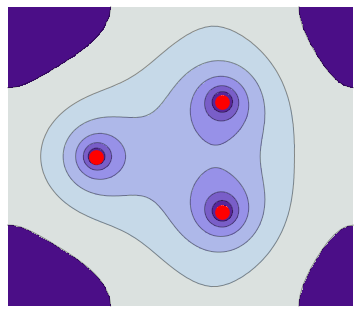} &
    \includegraphics[width=0.2\columnwidth]{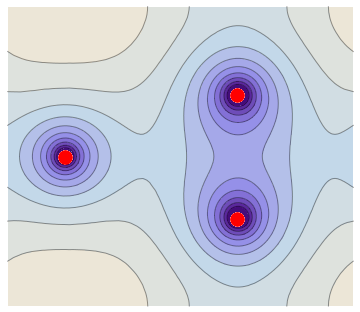}
    \tabularnewline
  \end{tabular}
  \caption{
    The spatial profile of CCS at $N_s=3$ for $\tau=\sqrt{\frac{3}{4}}\imath$ 
    where a) $w=0$, b) $w=\frac{1}{4}x_1$ and c) $w=\frac{1}{2}x_1$.
    These correspond to a rectangular lattice where we move the 
    $w$ away from $w=0$.
    No axes are drawn as each figure is one fundamental domain, $\Omega$.
    Notice how the spatial profile changes as $w$ is tuned away from $w=0$.
    This is a consequence of the zeros of $\varphi_w$ moving around.
    Upper:
    Contours of $|\varphi_w|^2$ with $\Omega$ centered at $z=w$.
    Brighter colours correspond to larger values of $|\varphi_w|^2$.
    Lower:
    Contours of $\log|\varphi_w|$ with $\Omega$ centered at $z=\frac 1 2(1+\tau)x_1$.
    A logarithmic scale is used to make the positions of the zeros (filled red circles) easier to see.
    \label{fig:spatial_profile_move_r_prime}}
\end{figure}

It is noteworthy that the $\sigma_x\sigma_y$-value of the CCS depends on the precise value of $w$.
If $w=x_n+\tau x_m$ the dispersion is at a minimum,
but if $w=x_{n+\frac{1}{2}}+\tau x_{m+\frac{1}{2}}$ the dispersion is at a maximum.
This suggest that the $x_1\times x_1\tau$ lattice is encoded in the CCS.
In the left panel of Figure \ref{fig:dxdx_of_Ns}, $\sigma_x\sigma_y$ is plotted for $\tau=\imath$ and $N_s=3$
where $w$ is varied over the fundamental domain.
A version of this is also seen in Figure \ref{fig:spatial_profile_move_r_prime} 
where in the upper panel the contours of $|\varphi_w|^2$ is shown for $w=0$,
$w=\frac{1}{4}x_1$ and $w=\frac{1}{2}x_1$ at $N_s=3$ and $\tau=\sqrt{\frac{3}{4}}\imath$.

For small values of $N_s$ the dispersion for $w\neq0$ can be higher than that of the corresponding LCS.
However, already at $N_s=9$ the maximum and minimum dispersions for CCS at different $w$ are practically indistinguishable. 
What happens at a more technical level is that the zeros of the wave function 
start moving around as can be seen in the lower panel of 
Figure \ref{fig:spatial_profile_move_r_prime}.
Here the torus region is held fixed, centered at $r=\frac{1}{2}(1+\tau)x_1$,
to make it easier to see that the sum of all the zeros is fixed 
as required by the boundary conditions in equation \eqref{eq:constraint}.

\begin{figure}
  \begin{tabular}{ccc}
    \includegraphics[width=0.4\columnwidth]{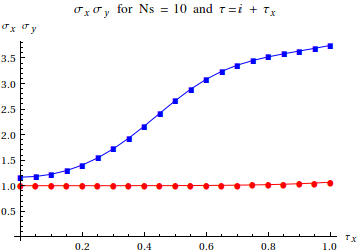} &
    \includegraphics[width=0.4\columnwidth]{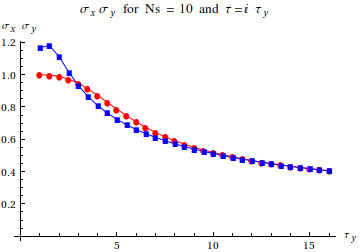} &
    \tabularnewline
    $a$ & $b$
  \end{tabular}
  \caption{
    The dispersion $\sigma_x\sigma_y$ for CCS (red curves) and LCS (blue curves) at $N_s=10$ as $\tau$ is varied.
    a) Here $\tau$ is varied from $\tau=\imath$ to $\tau=\imath+1$.
    As $\tau$ changes the CCS stay equally localized whereas the LCS becomes strongly distorted.
    b) Here $\tau$ is varied from $\tau=\imath$ to $\tau=16\imath$.
    As $\tau$ is changed there is a region where the LCS have slightly better dispersion.
    For large imaginary values of $\tau$ then $\sigma_x\sigma_y$ go to zero roughly as $\frac 1 {\sqrt{\Im(\tau)}}$,
    which is expected when $L_x$ is smaller than $\ell_B$ and the dispersion in $L_x$ direction is suppressed.
    \label{fig:dxdy_scan_tau}}
\end{figure}

Figure \ref{fig:dxdy_scan_tau} shows what happens when $\tau$ is tuned away from the square lattice at $\tau=\imath$.
We see here that there is a wide region of $\tau$ where the CCS have smaller $\sigma_x\sigma_y$ than the LCS.
However at large values of $\Im(\tau)$, corresponding to $L_y>L_x$,
the LCS seem to have a better dispersion than the CCS.
One should mention that this effect is tiny and starts to show first when the torus width
is so small that the CCS wave functions would be wider than the $L_x$ range.
If we instead change the real part of $\tau$ then the CCS functions become more localized in comparison with the LCS.
What happens with the LCS is that as $\tau\rightarrow\tau+\frac 1 2$ the torus geometry becomes triangular 
and the LCS develop two distinct maxima, as mentioned above.

\section{Holomorphic derivatives projected in the lowest landau level}\label{sec:Derivatives-in-LLL}

QH wave functions that describe hierarchical states on the plane and on the sphere,
are most simply expressed in terms of holomorphic derivatives acting on holomorphic polynomials.
In the composite fermion approach,
the derivatives are remnants of the anti holomorphic coordinates $z^\star$ present in higher LL wave functions\cite{Jain_07_Book}.
In the approach based on CFT,
they appear since the electron operators typically are Virasoro descendants of primary fields\cite{Soursa_11a}.
On the plane the derivatives are in principle easy to handle,
although it can be numerically difficult to take many derivatives of the polynomials,
which are of order $N^2$, where $N$ is the number of particles.
On the sphere the situation is more complicated because of the finite dimensionality of the space of LLL wave functions,
but recently the full hierarchy has been constructed also in this case\cite{Kvorning_11}.
On the torus the situation is more intricate.
As discussed in Ref. \onlinecite{Hermanns_08}, in the context of the CFT construction, 
the electrons are still most naturally represented by descendants of primary fields.
The corresponding conformal blocks, which are well defined on the torus, contain holomorphic derivatives.
Although these blocks do not satisfy the proper QH boundary conditions,
for the simple case of the Laughlin states, a subspace of functions that do can be found,
resulting in precisely the torus Laughlin wave functions, constructed by Haldane and Rezayi\cite{Haldane_85}.
For the hierarchical states, which on the plane involve holomorphic derivatives, this procedure fails.
A partial solution to this problem was proposed in Ref. \onlinecite{Hermanns_08},
where it was noticed that if the holomorphic derivative was replaced by a certain finite translation operator,
the wave functions were well defined, and had good overlap with the coulomb ground state, at least in certain geometries.

In this section we shall approach the problem from the point of view of coherent states,
and show that the occurrence of finite translations in Ref. \onlinecite{Hermanns_08} is not a coincidence.
The finite translations are related to how holomorphic derivatives of torus LLL wave functions, which are not themselves in the LLL, behave when projected back to the LLL.

\subsection{How to project holomorphic derivatives onto the LLL}
There are several routes to derive expressions for the projected derivatives.
The easiest way is to use ladder operators and cylindrical wave functions
and their overlaps as defined in Appendix \ref{sec:cyl-on-torus}.
It is important to note already now that the result will be sensitive to constant shifts in the gauge potential.
Constant shifts in the potential $A\rightarrow A+y^{\prime}$,
correspond to translations of the fundamental region.
An easy way to see that the choice of fundamental region matters is by considering $\phi(z)$ and $\phi(z+\tau L_y)$, where $\phi$ is a LLL wave function.
We see that $\partial_x\phi(z+\imath L_y)=-\imath L_ye^{-\imath xL_y}\phi(z)+e^{-\imath xL_y}\partial_x\phi(z)$.
Since $\partial_x\phi(z+\imath L_y) \neq \partial_x\phi(z)$ the following projection will not be the same.

We start by examining the effect of acting with one derivative on the basis states in the LLL.
Using the ladder operators we can express how the operators $y$, $\partial_y$ and  $\partial_x$ act on $\eta_{0,s}\equiv\eta_s$.
Since both $y$ and $\partial_x$ break the torus boundary conditions,
their actions are best described in terms of the cylinder functions $\chi_{n,s}$.
Letting $y$, $\partial_y$ and $\partial_x$  act on the basis functions $\eta_{ns}$ we get 
\begin{eqnarray*}
  y\eta_{ns} & = & \frac{1}{\sqrt{2}}(a+a^{\dagger})\eta_{ns}+\lambda_{ns}\\
  \partial_y\eta_{ns} & = & \frac{1}{\sqrt{2}}(a-a^{\dagger})\eta_{ns}\\
  \partial_x\eta_{ns} & = & -\imath\lambda_{ns}
\end{eqnarray*}
where $\lambda_{ns}=\sum_ty_{s+tN_s}a_{n,s+tN_s}\chi_{n,s+tN_s}$.
The coefficients $a_{n,s}$ were defined in Section \ref{sec:The-torus}.
Note that that $\lambda_{ns}$ has components in all Landau levels, since $\chi_{n,s}$ is not in a single Landau level.
The holomorphic derivatives are obtained as 
\begin{equation}
  \partial_z \eta_{ns}=\frac{1}{2\imath}\left(\lambda_{ns}+\frac{1}{\sqrt{2}}(a-a^{\dagger})\eta_{ns}\right).
  \label{eq:del_z_on_basis}
\end{equation}

As long as we are in the LLL then $\braOket{\eta_{0,s}}{\partial_z}{\eta_{0,r}}=\braOket{\eta_{0,s}}{\partial_{\bar{z}}}{\eta_{0,r}}$,
since $\braOket{\eta_{0,s}}{\partial_y}{\eta_{0,r}}=0$,
which means that to investigate the effect of $\partial_z$ we must calculate
$\braOket{\eta_{0,r}}{\partial_z}{\eta_{0,s}}=\frac{1}{2\imath}\braket{\eta_{0,r}}{\lambda_{0,s}}$.
For this purpose we need to know the overlap of the cylinder functions evaluated on the torus,
and this calculation is performed in Appendix B.
It is important that $\braOket{\eta_{0,s}}{\partial_z}{\eta_{0,r}}$ 
is proportional to $\delta_{r,s}$ with an $s$-dependent constant $G_s$.
The effect of projecting a derivative down to the LLL is
\begin{equation}
  \plll\partial_z\ket{\eta_{0,s}}=\frac{1}{2\imath}G_s\ket{\eta_{0,s}},
  \label{eq:PLLL_del_z_fi}
\end{equation}
so that $G_{s+N_s}=G_s$ and is thus periodic.
This enables us to write 
\begin{equation*}
  \plll\partial_z\ket{\eta_{0,s}}  = \frac{1}{2\imath}\sum_la_lt_1^l\ket{\eta_{0,s}}
\end{equation*}
where $a_l=\frac{1}{N_s}\sum_se^{\imath y_s x_l}G_s$ is the discrete Fourier transform of $G_s$.
Because the result does not depend on $s$ it holds for any state in the LLL,
and consequently 
\begin{equation}
  \plll\partial_z\phi(z)=\frac{1}{2\imath}\sum_la_l\phi(z+x_l)
  \label{eq:Derivative_projection}
\end{equation}
for any LLL state.

The same result can of course be obtained using LCS or CCS,
but the calculation is more complicated\footnote{
  In the LCS case one uses the resolution of unity
  to project a derivative on a LCS state as a weighted sum of all LCS states with coefficients $a_{mnm^{\prime}n^{\prime}}$.
  Using the self-reproducing kernel $\plcs$
  one can expand a generic state also in LCS states.
  After some algebra using the form of the weighted sum $a_{mnm^{\prime}n^{\prime}}$ one obtains the desired result.}.

\subsection{Higher order derivatives and derivatives in the thermodynamic limit}

In \eqref{eq:del_z_on_basis} we saw that $\partial_z\eta_{ns}$ will have components in all Landau levels because $\lambda_{ns}$.
It is therefore difficult to give a closed formula for $\plll\partial_z^k\eta_{ns}$.
We can however give a general formula for $(\plll\partial_z)^k\phi(z)$ since at each projection we can apply \eqref{eq:Derivative_projection}.
The effect of multiple derivatives on one state can thus be expressed as products of sums of translation operators,
which we can rewrite as a single sum of such operators:
\begin{equation*}
  (\plll\partial_z)^k  \sim  \sum_la_l^{(k)}t_x^l.
\end{equation*}
With the definition above, and $a_l^{(1)}\equiv a_l$,  we get a recursive relation 
\begin{equation}
  a_m^{(k)} = \sum_na_{m-n}^{(k-1)}a_n^{(1)} = \frac{1}{N_s}\sum_sG_s^ke^{\imath y_sx_m}
  \label{eq:c^(k)_m_of_G}
\end{equation}
where the coefficient $a_m^{(k)}$ is simply the $m$:th Fourier coefficients of $G_s^k$.
We cannot calculate the coefficient $a_l$ analytically, since it contains limited Gaussian integrals
(see \eqref{eq:G_s_re-rewritten} in Appendix \ref{sec:G_s}),
but we can take the thermodynamic limit where both $N_s$ and $L_y$ become large.
In this limit $\frac{s}{N_s}$ take almost continuous values and we can approximate $G_s$ with a continuous function 
\begin{equation}
  G(\xi) = \xi\sqrt{\pi}-\frac{1}{2}\sum_{k=-\infty}^{\infty}\int_{(k-\xi)L_y}^{(k+\xi)L_y}dy\, e^{-y^2}
  \label{eq:G_x}
\end{equation}
such that $G_s = G(\frac{s}{N_s})\frac{L_y}{\sqrt{\pi}}$.
By construction $G(\xi)$ has periodicity 1,
where we specialize to a torus centered around $z=0$.
We can simplify the expression for $G(\xi)$ by first taking a $\partial_\xi$ derivative, 
then expanding in Fourier modes and finally integrating in $\xi$.
The result is
\begin{equation*}
G(\xi) = \sum_{n=1}^{\infty}\frac{(-1)^ne^{-\frac{\pi^2n^2}{L_y^2}}}{\sqrt{\pi}n}\sin(2\pi n\xi),
\end{equation*}
and the Fourier coefficients of this function are
\begin{equation}
  a_n=\frac{(-1)^ne^{-\frac{\pi^2n^2}{L_y^2}}}{\sqrt{\pi}n}.
  \label{eq:c_n_thermodynamic}
\end{equation}
The $n=1$ term will dominate in the $L_y\rightarrow 0$ limit where $G(\xi)$ becomes a sine function,
which as $L_y\rightarrow\infty$ is changed to a sawtooth function.
At first this gives the impression that the dominance of the first term
is diminished in the thermodynamic limit.
This is certainly true but of little consequence since the number of terms 
that are reasonably large scales only as $n_{\mathrm{large}}\sim\sqrt{N_s}$.
The relative number of translations thus decreases as 
$\frac{n_{\mathrm{large}}}{N_s}\sim\frac{1}{\sqrt{N_s}}$ so that in the thermodynamic limit 
the relative number of the relevant translations goes down.
This is actually what one would expect since in the LLL,
possition cannot be specified closer than one magnetic length\footnote{
  One example of this is that on the plane $\varphi_w(z)$ is proportional to $e^{-|w-z|^2}$.
  Since the length of a $t_1$ translation goes as $\frac{L_x}{N_s}\sim\frac{1}{\sqrt{N_s}}$,
  the number of translations it would take to translate one magnetic length $\ell_B$ is
  $n_{\mathrm{trans}}\sim\sqrt{N_s}$.
  Thus the relative number of translations needed to move one magnetic
  length compared with the full width of the torus is $\frac{n_{\mathrm{trans}}}{N_s}\sim\frac{1}{\sqrt{N_s}}$.}.
  
\subsection{Holomorphic derivatives and  LLL wave functions}

We now return to the question of how to treat the holomorphic derivatives in the LLL CFT wave functions.
The problem we face is that from CFT we obtain wave functions that naturally contain derivative factors $\partial_z$.
Since these derivatives violate the boundary conditions and force the CFT wave function into higher Landau levels,
we would like to be able to project the $\partial_z$ onto the LLL in a consistent manner.
We have above found that under projection to the LLL a holomorphic derivative transforms as
$\partial_z \rightarrow \mathcal D\equiv\frac 1 {2\imath}\sum a_l t_x^l$,
where $a_l$ where given by \eqref{eq:c^(k)_m_of_G}.
This suggests that the substitution $\partial_z \rightarrow \mathcal D$ in the relevant conformal blocks
would yield good  hierarchical QH  wave functions on the torus.
This was in fact proposed in Ref. \onlinecite{Hermanns_08} but without any motivation,
and with the coefficients $a_l$ as free parameters.
This simple prescription will however not work for the following reason.
The quantum numbers for the many-particle wave functions can be taken as, 
$T_1\psi_j(z_1,\ldots,z_{N_e})=e^{\imath 2\pi \frac {K_j}{N_s}}\psi_j(z_1,\ldots,z_{N_e})$ and
$T_2\psi_j(z_1,\ldots,z_{N_e})=\psi_{j+1}(z_1,\ldots,z_{N_e})$, 
where $T_1=\prod_k t_{1,k}$ and $T_2=\prod_k t_{2,k}$ translates all coordinates one $t_1$-step or $t_2$-step respectively.
It is straight-forward to see that $\mathcal D$ does not commute with neither $T_2$ nor $T_2^q$,
and consequently does not respect the quantum numbers.
What will work, is the substitution,
\begin{equation}
  \prod_j \partial_j \rightarrow \sum_k  ( a_k  T_1^k + b_k T_2^k ),
  \label{eq:derivative_substitution}
\end{equation}
and it was in fact an ansatz of this form that, for reasons of simplicity, was tested numerically in  Ref. \onlinecite{Hermanns_08}. 
Since the simple substitution, $\partial_z \rightarrow \mathcal D$ does not work, it would be desirable to have 
some other guiding principle for finding the coefficients $a_k$ and $b_k$ in the expression \eqref{eq:derivative_substitution},
and it turns out that modular invariance is very important\cite{Hansson_12}. 

We now turn to the remark concerning the lattice coherent states. 
In Section \ref{sec:LCS} we defined the map $\plcs$ from the whole space of torus wave functions onto the LLL.
This map could be used to consistently bring the CFT wave functions downs to the LLL,
even where there are arbitrary functions of $\partial_z$.
The states that we have in mind are the ones that cannot be described 
in the hierarchy picture by simple condensations of quasi particles.
In these states anti-holomorphic components naturally arise as 
$\elliptic 1 {\bar z_i-\bar z_j} \tau$.
Transforming $\bar z\rightarrow\partial_z\rightarrow\mathcal D$ would be 
intractable given the amount of terms, but using $\plcs$ we could 
simply evaluate the trial wave function on the multidimensional lattice 
$\{z_{{m_1}{n_1}},\ldots,z_{{m_{N_e}}{n_{N_e}}}\}$ and construct the LLL counterpart from there.

The usage of $\plcs$ also opens up for an application related to calculating overlaps.
Given \eqref{eq:LCS_self_rep_kernel} we can calculate the overlap between two LLL states, $\psi$ and $\phi$ as 
\begin{equation*}
  \braket \phi \varphi = S^{-2} \sum_{m,m^\prime,n,n^\prime} 
  \phi^\star(z_{n^\prime m^\prime}) \varphi(z_{nm})
  \braket {m^\prime n^\prime} {mn},
\end{equation*}
where we here restrict ourselves to $\Re(\tau)=0$ for a cleaner notation.
For an $N_e$-body state $\varphi\left(z_{1},\ldots z_{N_e}\right)$, which can be expanded as 
\begin{equation*}
  \varphi\left(z_{1},\ldots,z_{N_e}\right)=S^{-N_e}\sum_{\mathbf{n},\mathbf{m}}\varphi\left(z_{n_{1}m_{1}},\ldots,z_{n_{N_e}m_{N_e}}\right)\prod_{j=1}^{N_e}\psi_{n_{j}m_{j}}\left(z_{j}\right),
\end{equation*}
the many-particle overlap is 
\begin{eqnarray*}
\braket{\phi}{\varphi} & = & 
S^{-2N_e}\sum_{\mathbf{n}^{\prime},\mathbf{m}^{\prime}}\sum_{\mathbf{n},\mathbf{m}}
\phi^{\star}\left(z_{n_{1}m_{1}},\ldots,z_{n_{N_e}m_{N_e}}\right)
\varphi\left(z_{n_{1}^{\prime}m_{1}^{\prime}},\ldots,z_{n_{N_e}^{\prime}m_{N_e}^{\prime}}\right)
\prod_{j=1}^{N_e}\braket{n_{j}^{\prime}m_{j}^{\prime}}{n_{j}m_{j}}.
\end{eqnarray*}

As this stands we would need to evaluate $\phi$ and $\varphi$ at approximately $\frac{N_{s}^{2N_e}}{N_e!}$ 
points each and calculate the $\frac{N_s^{4N_e}}{\left(N_e!\right)^{2}}$ cross terms. 
If we are bold and assume that only the diagonal $\braket {nm} {nm}$ terms will contribute,
and all other will either be small or be averaged to zero,
we can use the approximation $\braket{n^{\prime}m^{\prime}}{nm}\approx\delta_{n,n^{\prime}}\delta_{m,m^{\prime}}$.
In this case we would get
\begin{eqnarray*}
  \braket{\phi}{\varphi} & = & S^{-2N_e}\sum_{\mathbf{n},\mathbf{m}}
  \phi^{\star}\left(z_{n_{1}m_{1}},\ldots,z_{n_{N_e}m_{N_e}}\right)
  \varphi\left(z_{n_{1}m_{1}},\ldots,z_{n_{N_e}m_{N_e}}\right).
\end{eqnarray*}
which gives us roughly $\frac{N_s^{2N_e}}{N_e!}$ points to evaluate.
This is still a large number but we can here hopefully use that the QH system is strongly correlated,
to remove the (majority) of terms that are almost zero because two electron coordinates will be close to each other.
This could be used, not only as a systematic way of truncating an overlap calculation in the LLL,
but also to lessen the computational burden in using $\plcs$ to map 
higher Landau level wave functions down to the lowest one.


\section{Using CCS to project to the lowest Landau level}\label{sec:Laughlin4-1}

In this section we give an example of how the torus coherent state kernel can be used
to project many-body wave functions onto the lowest Landau level.
Following Girvin and Jach\cite{Girvin_84} we consider a short distance 
modification of the $\nu = \frac 1q$ Laughlin wave function  
\begin{equation}
  \tilde \Psi_{\frac 1q}  =  
  \plll e^{-\frac{q+2p}{4q}\sum_{j}\left|z_{j}\right|^{4}}\prod_{i<j}\left(z_{i}-z_{j}\right)^{q+p}\left(\bar{z}_{i}-\bar{z}_{j}\right)^{p}
  \propto e^{-\frac{1}{4}\sum_{j}\left|z_{j}\right|^{4}}\prod_{i<j}\left(\partial_{z_{i}}-\partial_{z_{j}}\right)^{p}\left(z_{i}-z_{j}\right)^{q+p},
  \label{eq:laughlin_alt_plane}
  \end{equation}
which has both $z$ and $\bar{z}$ components but no external derivatives.
Note that due to the Landau level projection there is no simple plasma analogy,
but $\tilde \Psi_{\frac 1q}$ still has the same maximal angular momentum as the usual Laughlin state.
Since $\tilde \Psi_{\frac 1q}$ is only a short distance modification of Laughlin's wave function,
we expect that this state is in the same universality class as the Laughlin state,
and a good trial wave function.
In the CFT approach to the quantum Hall hierarchy,
given in Refs. \onlinecite{Soursa_11a} and \onlinecite{Soursa_11b},
the wave function in the coherent state basis is given by products of 
holomorphic and anti holomorphic blocks as in \eqref{eq:laughlin_alt_plane}.
The wave function in this basis is calculated as the correlator of the electron operator $V(z)=e^{\imath\sqrt{q+p}\varphi_1(z)+\imath\sqrt{p}\bar\varphi_2(\bar z)}$.
The  electronic wave functions is obtained by projecting on the lowest Landau level,
which amounts to a convolution with a coherent state kernel. 
On the torus,
the latter point of view is fruitful since the coherent state wave function,
which is comparatively simple,
obeys the same boundary conditions as the much more complicated projected electron wave functions.
This is apparent since $t(\alpha)$ commutes with $\plll$ which is 
a convolution with the coherent state kernel, $\int d^2\xi\,\varphi_\xi(z)\,\psi(\xi)$.
From this it follows that the same periodic boundary conditions that apply for the coherent state wave function $t(L_x)\psi(z)=e^{\imath\phi}\psi(z)$,
also apply for the LLL projection, \emph{i.e.} $t(L_x)\plll\psi(z)=e^{\imath\phi}\plll\psi(z)$.

On the plane we can factor the electron operator into a chiral and an anti chiral sector and directly evaluate a correlator to get  \eqref{eq:laughlin_alt_plane}.
The situation on the torus is more complicated since there are several chiral and anti-chiral sectors,
and there is no procedure to factor the operators to directly get the trial wave functions.
Instead we follow Ref. \onlinecite{Hermanns_08} and use the transformation properties 
of the different conformal blocks to arrive at an appropriate subset of the full correlator.
The torus counterpart of the wave function \eqref{eq:laughlin_alt_plane} can be written 
as a single correlator of the full electron operator  $V(z)=e^{\imath\sqrt{q+p}\varphi_1(z,\bar z )+\imath\sqrt p\varphi_2(z,\bar z)}$,
where $\varphi_1$, $\varphi_2$ are compactified massless scalar fields with radii $R_1=\frac q{\sqrt{q+p}}$ and $R_2=\frac q{\sqrt{p}}$.

For notation, and the technical procedure for constructing  hierarchy wave functions on the torus, 
we refer to Ref. \onlinecite{Hermanns_08} and only quote the result here.
The full correlator can be written as a sum over conformal blocks
$\left\langle \prod_{i=1}^{N_{e}}V(z_i)\mathcal O_{\mathrm{bg}}\right\rangle=N\left(\tau\right)
\sum_{E_1,E_2}\psi_{E_1,E_2}\bar{\psi}_{\bar E_1,\bar E_2}$  where 
\begin{equation}
  \psi_{E_1,E_2}=e^{-\frac{q+2p}{2q}\sum_{i}y_{i}^{2}} 
  \prod_{i<j}\elliptic 1{z_{ij}}{\tau}^{q+p}\elliptic 1{-\bar{z}_{ij}}{-\bar{\tau}}^{p}
  \mathcal{F}_{E_1,E_2}\left(Z|\tau\right)
\end{equation}
and $N(\tau)$ is a normalization that will be of no importance here. Here $Z=\frac{1}{L_x}\sum_kz_k$ and $z_{ij}=\frac 1{L_x}(z_i-z_j)$.
The sum over $E_1$ and $E_2$ runs over a lattice spanned by 
$E_j=\frac {e_j}{R_j}+\frac{m_j R_j}2$ and  $\bar E_j=\frac {e_j}{R_j}-\frac{m_j R_j}2$ 
where $e_j$ and $m_j$ are integers.
The conformal blocks are given by
\begin{equation}
  \mathcal{F}_{E_1,E_2}\left(Z|\tau\right)=e^{\imath\pi\left[\tau E_1^2-\bar{\tau}E_2^2\right]}e^{2\pi\imath\left[E_1\sqrt{q+p}Z-E_2\sqrt p\bar Z\right]}.
\end{equation}
Fixing the boundary conditions in the $L_x$ and $\tau L_x$ directions amounts to selecting a subset of the $E_1$, $E_2$ lattice.
This subset is parametrized by $(E_1,E_2)=k(\sqrt{q+p},\sqrt p)+\Gamma_0$ where $k\in\mathbb Z$.
The simplest choice of the offset $\Gamma_0=r(\sqrt{q+p},\sqrt p)$ gives the center of mass function
\begin{equation}
  \label{eq:Laughlin_H_r}
  \mathcal{H}_{r}=\sum_k \mathcal{F}_{(r+k)\sqrt{q+p},(r+k)\sqrt p}=\ellipticgeneralized r0{\left(q+p\right)Z-p\bar Z}{\tau\left(q+p\right)-\bar{\tau}p}.
\end{equation} 
Here $\ellipticgeneralized abz\tau$ is a generalized $\vartheta$-function such that $\mathcal H_{r+1}=\mathcal H_r$.
Under center of mass translations $T_2=\prod_k t_{2,k}$ we find that $\mathcal H_r\rightarrow\mathcal H_{r+\frac 1q}$
and we have recovered the expected $q$-fold degeneracy of a $\nu=\frac 1q$  state.
The center of mass piece \eqref{eq:Laughlin_H_r} implies one-particle periodic boundary conditions with phases $\phi_1=\pi(N_e-1)+2\pi qr$ and $\phi_2=\pi(N_e-1)$.
We can select arbitrary boundary conditions $\phi_1$ and $\phi_2$ by translating the center of mass coordinate $Z$ appropriately.
After such a magnetic translation the final modified Laughlin wave function for 
$\phi_1=\phi_2=0$ is given by
\begin{equation}
  \psi_n^{(q,p)}=e^{-\frac{q+2p}{2q}\sum_{i}y_{i}^{2}}\prod_{i<j}
  \elliptic 1{z_{ij}}{\tau}^{q+p}\elliptic 1{-\bar{z}_{ij}}{-\bar{\tau}}^{p}
  \ellipticgeneralized{\frac nq+\alpha}{\alpha}{\left(q+p\right)Z-p\bar Z}{\tau\left(q+p\right)-\bar{\tau}p}
  \label{eq:Mod_Laughlin_final}
\end{equation}
where $n=1,\ldots,q$ enumerates the different momentum sectors
and $\alpha=\frac 12(N_e-1)$.
Furthermore, since we already have showed that the boundary conditions are preserved under projection onto the LLL,
we have constructed a torus version of the wave function \eqref{eq:laughlin_alt_plane}.

We end this section by numerically comparing \eqref{eq:Mod_Laughlin_final} with exact diagonalization of the Coulomb potential at $\tau=\imath$.
In their original work Girvin and Jach noted that the Laughlin state 
$\psi^{(q,0)}$ could be improved by considering components with $p\neq 0$,
and on the torus we observe the same thing.
For $p\neq 0$ then $\psi^{(q,p)}$ is not entirely in the LLL,
but the projected wave functions $\plll\psi^{(q,p)}$ do still have 
good overlaps with the Coulomb ground state.
For $N_e=3$ electrons and $N=3\times 10^6$ Monte Carlo points, the $(q,p)=(3,2)$ state has the best overlap with exact Coulomb,
$|\braket {\psi_{\mathrm{Coulomb}}} {\psi^{(3,2)}} |^2=0.9999(4\pm6)$\footnote{
  The uncertainty is estimated by partitioning the Monte Carlo points into
  $n$ sets and redoing the analysis with the partitions.
  The error is computed as the standard deviation divided by $\sqrt n$.}.
This should be compared to Laughlin, which has 
$|\braket {\psi_{\mathrm{Coulomb}}} {\psi_{\mathrm{Laughlin}}} |^2=0.9990(0\pm2)$.
For $N_e=4$ electrons and $N=3\times 10^7$ Monte Carlo points,
the $(q,p)=(3,1)$ state matches Coulomb best,
with $|\braket {\psi_{\mathrm{Coulomb}}} {\psi^{(3,1)}} |^2=0.9976(5\pm6)$
compared to $|\braket {\psi_{\mathrm{Coulomb}}} {\psi_{\mathrm{Laughlin}}} |^2=0.9792(8\pm3)$
for the Laughlin state.
Going to larger system is exponentially difficult as the projection on the LLL requires an overlap calculation with all basis states.


\section{Summary}

In this paper we studied two alternative ways of realizing 
coherent states in the lowest Landau level on a torus.
We explored the set of lattice coherent state wave functions and found 
that they can be used to define a map $\plcs$ which for the two lowest 
Landau levels acts like the true projection $\plll$.

We also examined the continuous coherent state wave functions,
obtained by $\plll\delta(z-z^{\prime})$,
and found that these indeed seem to minimize the spatial dispersion.
The CCS turn out not to be entirely homogeneous with respect to the
guiding center position $w$ but rather periodic with periods
dictated by the lattice of broken translations $x_1\times\tau x_1.$

We considered the effect of projecting a holomorphic derivative to the LLL.
There we found that the projected derivative becomes a sum of translation operators $\plll\partial_z\sim\mathcal D=\sum_l a_lt_x^l$.
The projection could be used to systematically transform derivatives to $\mathcal D$ but is computationally expensive,
and does not necessarily respect modular invariance.

We also proposed a map $\plcs$ that could be used instead of,
or in conjunction with, 
$\partial_z\rightarrow\mathcal D$ to map wave functions to the LLL.
For simple hierarchical states,
like those for $\nu=\frac{2}{5}$ where there are only particle condensates,
the many body states already reside in the LLL and so $\plcs$ is redundant.
For more complicated states,
like $\nu=\frac{2}{3}$ where there are hole condensates,
there naturally arise anti-holomorphic components of the trial wave functions.
Here usage of $\plcs$ could be one way of mapping these functions to the lowest Landau level.

Finally we considered an explicit example of states that have anti-holomorphic components,
constructed trial wave functions for them in the coherent state basis
and compared them numerically with the ground state of the coulomb potential.

\begin{acknowledgments}
  I would like to thank my supervisor Thors Hans Hansson for guidance and Maria Hermanns and Juha Suorsa for discussions.
  A special thanks to Thomas Kvorning who pointed out an error in the original
  version of the manuscript.
  This work was supported by the Swedish Science Research Council.
\end{acknowledgments}

\appendix


\section{Orthonormal basis states on arbitrary $\tau$ torus}\label{sec:Orthonormal-basis-states} 

We will here summarize some information on the basis functions that exist on the torus.
These can be built from the cylinder functions under the constraints that 
\begin{eqnarray}
  t_1^l\eta_{n,s} & = & e^{-\imath y_lx_s}\eta_{n,s}\label{eq:app:t1_phi}\\
  t_2\eta_{n,s} & = & \eta_{n,s-k}\label{eq:app:t2_phi}.
\end{eqnarray}
As in the main text we parametrize our torus as $L_x\times L_x\tau$ where $L_x\tau=L_\Delta+\imath L_y$.
Thus we think of the height of the parallelogram as $L_y$ and the skewness as $L_\Delta$.
The flux relation is as usual $L_xL_y=2\pi N_s$.
The cylinder wave functions in the $n$:th Landau level have the form
\begin{equation}
  \chi_{n,s} = \frac{1}{\sqrt{L_x\sqrt{\pi}}}e^{-\imath y_sx}H_n(y-y_s)e^{-\frac{1}{2}(y-y_s)^2}.
  \label{eq:cylinder_higher_landau_levels}
\end{equation}
Under small magnetic translations they obey
\begin{eqnarray*}
  t_1^k\chi_{n,s} & = & e^{-\imath y_sx_k}\chi_{n,k}\\
  t_2^k\chi_{n,s} & = & e^{-\imath y_s\omega_k}e^{\imath\frac{y_k\omega_k}{2}}\chi_{n,s-k}.
\end{eqnarray*}

We see that to fulfill \eqref{eq:app:t1_phi} we must construct 
$\eta_{n,s}=\sum_t\alpha_{s,t}\chi_{n,s+t N_s}$ where $\alpha_{s,t}$ is a weight.
To fulfill \eqref{eq:app:t2_phi} we need to choose $\alpha_{s,t}$ to be a phase.
The linear combination needed has the form $\eta_{n,s}=\sum_te^{\imath\beta_{s+tN_s}}\chi_{n,s+tN_s}$.
We now have that 
\begin{equation*}
  t_2^k\eta_{n,s}=\sum_te^{\imath\beta_{s+tN_s}}e^{-\imath y_{s+tN_s}\omega_k}
  e^{\imath\frac{y_k\omega_k}{2}}\chi_{n,s+tN_s-k}
\end{equation*}
and want to choose $\beta_{s+tN_s}$ such that $\beta_{s+tN_s}-y_{s+tN_s}\omega_k+\frac{y_k\omega_k}{2}=\beta_{s+tN_s-k}$ modulo $2\pi$.
We see that this is fulfilled if we choose $\beta_r=\frac{1}{2}y_r\omega_r$ such that 
\begin{equation*}
  \eta_{n,s}=\sum_l\delta_{s,l}^{(N_s)}e^{\imath\frac{1}{2}y_l\omega_l}\chi_{n,l}=\sum_te^{\imath\frac{1}{2}(y_s+tL_y)(\omega_s+tL_\Delta)}\chi_{n,s+tN_s}.
\end{equation*}

The full wave functions in the LLL are thus
\begin{equation}
  \eta_s(z)  =  \frac{1}{\sqrt{L_x\sqrt{\pi}}}\sum_te^{\imath\frac{1}{2}(y_s+tL_y)(\omega_s+tL_\Delta)}e^{-\imath(y_s+tL_y)x}e^{-\frac{1}{2}(y-y_s-tL_y)^2}
  \label{eq:phi_expanded_raw}
\end{equation}
which we can, up to a factor of $e^{-\imath\frac{1}{2}y_s\omega_s}$, rewrite as
\begin{equation*}
  \eta_s(z)=\frac{e^{-\frac{1}{2}(y-y_s)^2}}{\sqrt{L_x\sqrt{\pi}}}e^{-\imath y_sx}\elliptic 3{\frac{\pi N_s}{L_x}(z-\tau x_s)}{N_s\tau}.
\end{equation*}


\section{Cylinder functions on the torus}\label{sec:cyl-on-torus}
Let us evaluate the overlap between the cylinder wave functions from \eqref{eq:cylinder_functions} taken over the torus geometry.
Normally one would not look at this since these functions do not form an orthonormal set on the torus.
In our case we will be interested in this overlap, since when $\partial_z$ acts on periodized functions it will spoil the boundary conditions.
To keep things general we will integrate over a torus resting with its lower left corner at a point $z=\alpha+\imath\beta$.
The overlap, which is $\alpha$ dependent, is
\begin{equation}
  \braket{\chi_{mr}}{\chi_{ns}}_{\mathrm{torus}}  =  \delta_{rs}I_s^{m,n}
  \label{eq:cylinder_overlap_torus}
\end{equation}
where 
\begin{equation}
  I_s^{m,n}=\frac{1}{\sqrt{\pi}}\int_{\beta-y_s}^{\beta-y_s+L_y}dy\, H_m(y)H_n(y)e^{-y^2}.
  \label{eq:cylinder-overlap-torus}
\end{equation}
The integral has the property that $\sum_tI_{s+tN_s}^{m,n}=\delta_{mn}$.
With the knowledge of the overlap of cylinder wave functions on the torus we see that 
the $\eta_{n,s}$ form an orthonormal set of states
\begin{equation*}
\braket{\eta_{mr}}{\eta_{ns}}=\sum_{t,l}e^{\imath\alpha_{r+tN_s}-\imath\alpha_{s+lN_s}}\braket{\chi_{m,r+tN_s}}{\chi_{n,s+lN_s}}_{\mathrm{torus}}=\delta_{r,s}\delta_{m,n}.
\end{equation*}
From this follows that $\mathcal{P}_m=\sum_r\ketbra{\eta_{mr}}{\eta_{mr}}$ 
is a projector onto the $m$:th Landau level and 
$\mathbb{I}=\sum_m\mathcal{P}_m$ is the identity operator.

The $\eta_{n,s}$ were given by 
$\eta_{n,s}=\sum_r\delta_{r,s}^{(N_s)}e^{\imath\frac{1}{2}y_r\omega_r}\chi_{n,r}$.
Remembering that $\partial_x\chi_{n,s}=-\imath y_s\chi_{n,s}$ we see that we can write 
a $\partial_x$ derivative on $\eta_{ns}$ as 
\begin{eqnarray*}
  \partial_x\eta_{n,s} & = & -\imath\sum_t
  e^{\imath\frac{1}{2}y_{s+tN_s}\omega_{s+tN_s}}(y_s+tL_y)\chi_{n,s+tN_s}\\
  & = & -\imath y_s\eta_{n,s}-\imath L_y\sum_tt
  e^{\imath\frac{1}{2}y_{s+tN_s}\omega_{s+tN_s}}\chi_{n,s+tN_s}.
\end{eqnarray*}
We also see that if we want to calculate the overlap of $\braket{\eta_{m,r}}{\partial_x\eta_{n,s}}$ we end up calculating 
\begin{eqnarray*}
  \braket{\eta_{m,r}}{\partial_x\eta_{n,s}} & = & 
  -\imath y_s\braket{\eta_{m,r}}{\eta_{n,s}}-
  \imath L_y\braket{\eta_{m,r}}{\sum_tte^{\imath\frac{1}{2}y_{s+tN_s}\omega_{s+tN_s}}\chi_{n,s+tN_s}}\\
  & = & -\imath y_s\delta_{m,n}\delta_{r,s}-
  \imath L_y\sum_{l,t}e^{-\imath\frac{1}{2}y_{r+lN_s}\omega_{r+lN_s}}t
  e^{\imath\frac{1}{2}y_{s+tN_s}\omega_{s+tN_s}}\braket{\chi_{m,r+lN_s}}{\chi_{n,s+tN_s}}_{\mathrm{torus}}\\
  & = & -\imath\delta_{r,s}\left(y_s\delta_{m,n}+L_y\sum_ttI_{s+tN_s}^{m,n}\right).
\end{eqnarray*}
It is the part inside the parentheses that we will call $G_s^{m,n}=y_s\delta_{m,n}+L_y\sum_ttI_{s+tN_s}^{m,n}$.
We see by inspection that $G_s^{m,n}$ is periodic with $G_s^{m,n}=G_{s+N_s}^{m,n}$.


\section{Rewriting $G_s$}\label{sec:G_s}
If we are looking at functions in the LLL we can simplify the expression of $G_s=G_s^{00}$.
To facilitate further manipulations we will rewrite $G_s$.
The main difficulty comes from $\sum_ttI_{s+tN_s}^{00}$ which can be rewritten as 

\begin{eqnarray*}
  \sqrt{\pi}\sum_ttI_{s+tN_s}^{00} & = & \sum_tt\int_{\alpha+tL_y}^{\alpha+(t+1)L_y}dy\, e^{-y^2}\\
  & = & -\frac{1}{2}\sqrt{\pi}-\frac{1}{2}\sum_t\int_{tL_y-\alpha}^{tL_y+\alpha}dy\, e^{-y^2}
\end{eqnarray*}
where $\alpha=\beta-y_s$. This gives $G_s$ as
\begin{equation}
  G_s=y_s-\frac{L_y}{2}+
  \frac{L_y}{2\sqrt{\pi}}\sum_t\int_{tL_y+\beta-y_s}^{tL_y-\beta+y_s}dy\, e^{-y^2}
  \label{eq:G_s_rewritten}
\end{equation}
which is numerically simpler since we are only integrating over patches of Gaussian functions.
Reparametrizing $\beta=-\frac{L_y}{2}+\delta$,
such that $\delta$ is the deviation from the symmetric integration region we have
\begin{equation}
  G_s=y_s-\frac{L_y}{2\sqrt{\pi}}\sum_t\int_{tL_y+\delta-y_s}^{tL_y-\delta+y_s}dy\, e^{-y^2}.
  \label{eq:G_s_re-rewritten}
\end{equation}


\bibliographystyle{unsrt}
\bibliography{CSW_ref}

\end{document}